\documentclass[useAMS,usenatbib]{mn2e}
\usepackage{Times}
\usepackage{psfrag}
\usepackage{pspicture}
\usepackage{graphicx}

\title[The clustering and evolution of H$\alpha$ emitters at $\bf z\sim1$ from HiZELS]{The clustering and evolution of H$\alpha$ emitters at $\bf z\sim1$ from HiZELS\thanks{This work is based on observations obtained using the Wide Field CAmera (WFCAM) on the 3.8m United Kingdom Infrared Telescope (UKIRT), as part of the Hi-$z$ Emission Line Survey (HiZELS).} }
\author[D. Sobral et al.]{David Sobral$^{1}$\thanks{E-mail: drss@roe.ac.uk}, Philip N. Best$^{1}$, James E. Geach$^{2}$, Ian Smail$^{2}$, Michele Cirasuolo$^{3}$,
\newauthor Timothy Garn$^{1}$, Gavin B. Dalton$^{4,5}$ and Jaron Kurk$^{6}$\\
$^{1}$SUPA, Institute for Astronomy, Royal Observatory of Edinburgh, Blackford Hill, Edinburgh, EH9 3HJ, UK\\
$^{2}$Institute for Computational Cosmology, Durham University, South Road, Durham, DH1 3LE, UK\\
$^{3}$Astronomy Technology Centre, Royal Observatory of Edinburgh, Blackford Hill, Edinburgh, EH9 3HJ, UK\\
$^{4}$Astrophysics, Department of Physics, Keble Road, Oxford, OX1 3RH, UK\\
$^{5}$Space Science and Technology, Rutherford Appleton Laboratory, HSIC, Didcot, OX11 0QX, UK\\
$^{6}$Max-Planck-Institut f{\"u}r Extraterrestrische Physik, Postfach 1312, 85741 Garching, Germany\\
}
\begin{document}

\date{Accepted 2010 January 18. Received 2010 January 15; in original form 2009 December 21}

\pagerange{\pageref{firstpage}--\pageref{lastpage}} \pubyear{2009}

\maketitle
\label{firstpage}
\begin{abstract}

The clustering properties of a well-defined sample of 734 H$\alpha$ emitters at $z=0.845\pm0.015$, obtained as part of the Hi-$z$ Emission Line Survey (HiZELS), are investigated. The spatial correlation function of these H$\alpha$ emitters is very well-described by the power law $\xi=(\rm r/r_0)^{-1.8}$, with a real-space correlation, r$_0$, of $2.7 \pm 0.3$\,h$^{-1}$\,Mpc. The correlation length r$_0$ increases strongly with H$\alpha$ luminosity (L$_{\rm H\alpha}$) , from r$_0\sim2$\,h$^{-1}$\,Mpc for the most quiescent galaxies (star-formation rates of $\sim4$\,M$_{\odot}$yr$^{-1}$), up to r$_0>5$\,h$^{-1}$\,Mpc for the brightest galaxies in H$\alpha$. The correlation length also increases with increasing rest-frame $K$-band ($M_K$) luminosity, but the r$_0$-L$_{\rm H\alpha}$ correlation maintains its full statistical significance at fixed $M_K$. At $z=0.84$, star-forming galaxies classified as irregulars or mergers are much more clustered than discs and non-mergers, respectively; however, once the samples are matched in L$_{\rm H\alpha}$ and $M_K$, the differences vanish, implying that the clustering is independent of morphological type at $z\sim1$ just as in the local Universe. The typical H$\alpha$ emitters found at $z=0.84$ reside in dark-matter haloes of $\approx10^{12}$\,M$_{\odot}$, but those with the highest SFRs reside in more massive haloes of $\approx10^{13}$\,M$_{\odot}$. The results are compared with those of H$\alpha$ surveys at different redshifts: although the break of the H$\alpha$ luminosity function $L_{\rm H\alpha}^*$ evolves by a factor of $\sim30$ from $z=0.24$ to $z=2.23$, if the H$\alpha$ luminosities at each redshift are scaled by $L_{\rm H\alpha}^*(z)$ then the correlation lengths indicate that, independently of cosmic time, galaxies with the same (L$_{\rm H\alpha}$)/$L_{\rm H\alpha}^*(z)$ are found in dark matter haloes of similar masses. This not only confirms that the star-formation efficiency in high redshift haloes is higher than locally, but also suggests a fundamental connection between the strong negative evolution of $L_{\rm H\alpha}^*$ since $z=2.23$ and the quenching of star-formation in galaxies residing within dark-matter haloes significantly more massive than $10^{12}$\,M$_{\odot}$ at any given epoch.

\end{abstract}

\section{Introduction}\label{intro}

In a Universe dominated by cold dark matter, galaxies are found in dark matter haloes with a structure determined by universal scaling relations \citep[e.g.][]{Navarro1996}. Since baryons trace the underlying distribution of dark matter, measurements of the clustering of baryonic matter can be used to extract typical dark-matter halo masses \citep{Mo1996,Sheth2001} and to suggest links between populations found at different epochs.

Wide surveys of the nearby Universe \citep[e.g. 2dF, SDSS;][]{colless,sloan} have now assembled extremely large samples of galaxies which can be explored to study their clustering properties in great detail. Using those data, several studies have found that the amplitude of the two-point correlation function rises continuously with galaxy luminosity \citep[e.g.][]{Norberg2001,Zehavi,Liii}. They also reveal that the most rapid increase occurs above the characteristic luminosity, $L^*$. Red galaxies are found to be more clustered than blue galaxies, but the correlation amplitude of the latter population also increases continuously with blue or near-infrared luminosities \citep[e.g.][]{Zehavi}. This seems to indicate that both star-formation rate (SFR, traced by ultraviolet/blue light) and stellar mass (traced by near-infrared light) are important for determining the clustering of different populations of galaxies. Other studies have used morphological classifications. \cite{Skibba}, for example, took advantage of the largest sample of visually classified morphologies to date (from SDSS) to clearly reveal that although early-type galaxies cluster more strongly than discs, once the analysis is done at a fixed colour and luminosity no significant difference is found; this confirms previous results \citep[e.g.][]{Beisbart, Ball} for the nearby Universe.

Understanding when these trends were created and how they evolved with cosmic time is a key input to galaxy formation models and to our general understanding of how galaxies formed and evolved. The first clustering studies of different populations of galaxies beyond the local Universe \citep[e.g.][]{Efstathiou,Fisher1994,Brainerd,LeFevre1996}, although pioneering, were mostly limited by the lack of information on the redshifts for their samples. Fortunately, those problems are now starting to be effectively tackled by larger and deeper surveys. Such surveys have recently led to robust clustering studies of specific populations of sources at moderate and high redshift such as AGN \citep[e.g.][]{Angela}, sub-mm galaxies \citep[e.g.][]{Blain04,Weiss}, luminous red and massive galaxies \citep[e.g.][]{wake}, or star-forming galaxies such as H$\alpha$ emitters, Lyman-break galaxies or Lyman-$\alpha$ emitters \citep[e.g.][]{G08,McLure2009,Shioya2009}. A dependence of the clustering  on galaxy luminosity has also been identified beyond the local Universe \citep[e.g.][]{Gianvalisco2001}. In particular, \cite{Kong}, \cite{Hayashi}  and, more recently, \cite{Hartley} found a clear dependence of the clustering amplitude on near-infrared luminosity for $z\sim1-2$ ``B$z$K'' selected galaxies, indicating that already in the young Universe the most massive galaxies were much more clustered than the least massive ones. These results mean that although different studies have been combined to suggest links between the same population at different redshifts, or between different populations at different epochs, interpreting these requires a great deal of care, since luminosity limits typically increase significantly with redshift and are different for different populations. 

Narrow-band H$\alpha$ surveys are now a very effective way to obtain representative samples of star-forming galaxies. Since the \cite{Gallego1995} wide-area study in the local Universe, tremendous progress has been achieved, with recent narrow-band studies \citep[e.g.][]{Ly2007,Shioya2008,G08,Shim09,S09} taking advantage of state-of-the-art wide-field cameras and obtaining large samples of typically hundreds of H$\alpha$ emitters from $z=0$ to $z=2.23$, down to limiting star-formation rates (SFRs) of $\approx$ 1-10 M$_{\odot}$yr$^{-1}$. In particular, HiZELS, the Hi-$z$ Emission Line Survey (c.f. Geach et al. 2008, Sobral et al. 2009a; hereafter G08 and S09, respectively), is obtaining and exploring unique samples of star-forming galaxies presenting H$\alpha$ emission \citep[and other major emission lines, e.g.][]{S09b}, redshifted into the $J$, $H$ and $K$ bands. By using a set of existing and custom-made narrow-band filters, HiZELS is surveying H$\alpha$ emitters at $z=0.84$, $z=1.47$ and $z=2.23$ over several square degree areas of extragalactic sky.

Narrow-band surveys have a great potential for determining the clustering properties of large samples of galaxies and their evolution with cosmic time. Such surveys probe remarkably thin redshift slices ($\Delta z\approx0.02$), which not only provides an undeniable advantage over photometric surveys that can be significantly affected by systematic uncertainties, but also allows the study of very well-defined cosmic epochs. Additionally, the selection function is well-understood and easy to model in detail, a feature that contrasts deeply with that of current large high-redshift spectroscopic surveys. Narrow-band surveys also populate the redshift slices they probe with high completeness down to a known flux limit, in contrast to spectroscopic surveys which are usually very incomplete at any single redshift and present the typical pencil-beam distribution problems. Finally, narrow-band surveys can select equivalent populations at different redshifts, making it possible to really understand the evolution with cosmic time, avoiding the biases arising from comparing potentially different populations.

\begin{figure*}
\centering
\includegraphics[width=16.2cm]{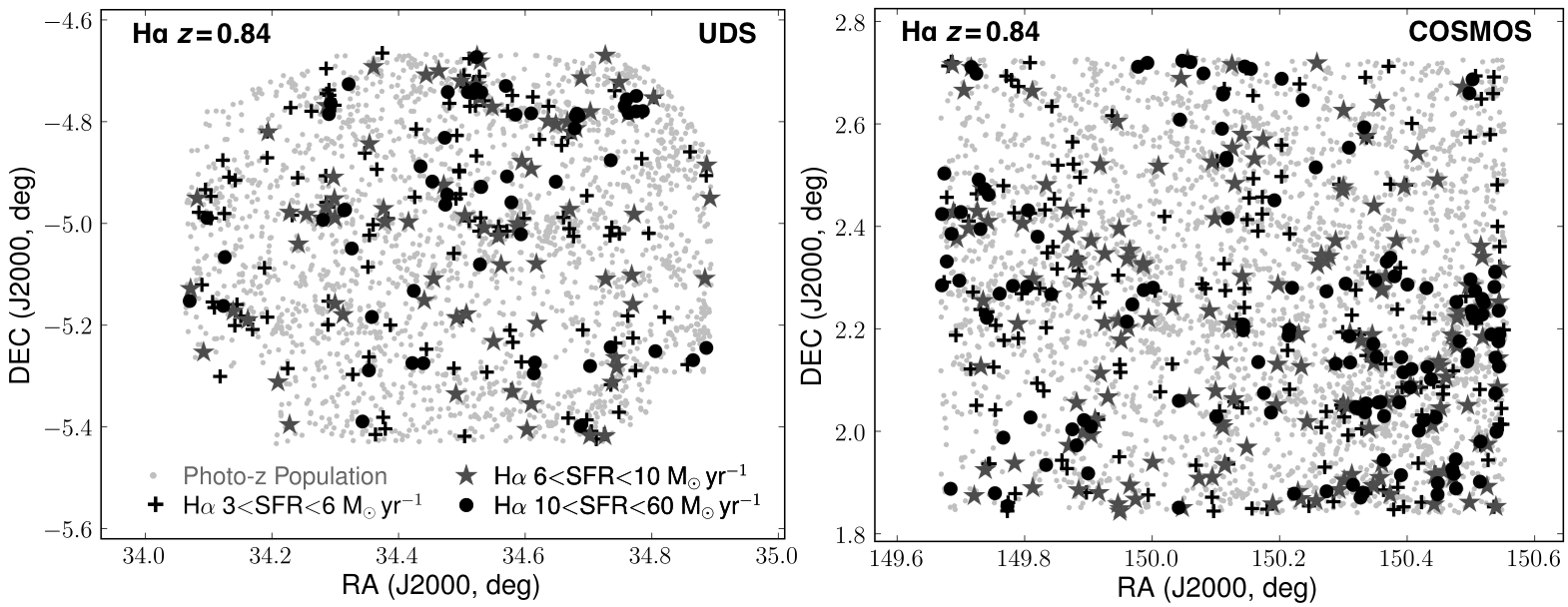}
\caption[on-sky COSMOS_UDS]{The on-sky distribution of the H$\alpha$ emitters found at $z=0.84$ in the UDS (left panel) and COSMOS (right panel) fields, together with a photometric redshift selected sample at the same redshift. Both panels/boxes cover the same angular/physical area, corresponding to $\approx29\times29$\,Mpc at $z=0.84$. The H$\alpha$ emitters are plotted in 3 different symbols corresponding to different star-formation rates (see key in left panel). \label{on_sky}}
\end{figure*}

This paper presents a detailed clustering study of the largest sample of narrow-band selected H$\alpha$ emitters at $z\sim1$ (c.f. S09), obtained after conducting deep narrow-band imaging in the $J$ band over $\sim 1.3\deg^2$, as part of the HiZELS survey. It is organised in the following way. \S2 outlines the data, the samples and their on-sky distribution. \S3 presents the angular two-point correlation function, carefully estimating the errors and other potential bias, along with the exact calculation of the real-space correlation. \S4 quantifies the clear clustering dependences on the host galaxy properties. \S5 presents the first detailed comparison between the clustering properties of H$\alpha$ emitters from $z=0.24$ up to $z=2.23$. Finally, \S6 presents the conclusions. An H$_0=100$h km\,s$^{-1}$\,Mpc$^{-1}$, $\Omega_M=0.3$ and $\Omega_{\Lambda}=0.7$ cosmology is used and, except where otherwise noted, magnitudes are presented in the Vega system and h $=0.7$.

\section{DATA AND SAMPLES}\label{data_technique}

\subsection{The sample of H$\alpha$ emitters from HiZELS}\label{sample}

This paper takes advantage of the HiZELS large sample of H$\alpha$ emitters at $z=0.845$ presented in S09 (the reader is referred to that paper for full details of how the sample was constructed). Briefly, the sample was obtained from data taken through a narrow-band $J$ filter ($\lambda_{\rm eff} = 1.211\pm0.015\umu$m) using the Wide Field CAMera (WFCAM) on the United Kingdom Infrared Telescope (UKIRT), and reaches an effective flux limit of $8\times10^{-17}$\,erg\,s$^{-1}$\,cm$^{-2}$ over $\sim 0.7\deg^2$ in the UKIDSS Ultra Deep Survey \citep[UDS,][]{2007MNRAS.379.1599L} field, and $\sim 0.8\deg^2$ in the Cosmological Evolution Survey \citep[COSMOS,][]{2007ApJS..172....1S} field. The data allowed the selection of a total of 1517 line emitters which are clearly detected in NB$_{\rm J}$ (signal-to-noise ratio $>3$) with a $J$-NB$_{\rm J}$ colour excess significance of $\Sigma>2.5$ and observed equivalent width EW$ >50$ \AA.

As the detected emission may originate from several possible emission lines at different redshifts, photometric redshifts\footnote{Photometric redshifts used in S09 for COSMOS present $\sigma(\Delta z ) = 0.03$, where $\Delta z$ = $(z_{phot}-z_{spec})/(1+z_{spec})$; the fraction of outliers, defined as sources with $\Delta  z>3\sigma(\Delta z)$, is lower than 3per cent, while for UDS, the photometric redshifts have $\sigma(\Delta z) = 0.04$, with 2 per cent of outliers.} were used to distinguish between them and select the complete sample of 743 H$\alpha$ emitters over a co-moving volume of $1.8 \times 10^5$\,Mpc$^3$ at $z=0.84$. Of these, 477 are found in COSMOS (0.76 deg$^2$) and 266 in UDS (0.54 deg$^2$; the reduction in area is driven by the overlap with the high-quality photometric catalogue used -- c.f. S09 for more details)). The completeness and reliability of this sample were studied using the $\sim10^4$ available redshifts from $z$COSMOS Data Release 2 \citep{zCOSMOS}, spectroscopically confirming $\sim100$ H$\alpha$ emitters within the photometric redshift selected sample; this allowed to estimate a $>95$ per cent reliability and $>96$ per cent completeness for the sample in COSMOS. H$\alpha$ fluxes, H$\alpha$ luminosities and H$\alpha$-based star-formation rates were obtained for all H$\alpha$ candidates using HiZELS data -- the details on how these were computed are fully detailed in S09.

Since the work presented in S09, improved photometric redshifts for COSMOS and UDS have been produced, incorporating a higher number of bands, deeper data and accounting for possible emission-line flux contamination of the broad-bands \citep[c.f.][Cirasuolo et al. in prep.]{Cirasuolo08,Ilbert09}; these further confirm the robustness and completeness of the selection done in S09. Nevertheless, the H$\alpha$ sample used for the analysis in this paper is slightly modified from that in S09 on the basis of the new photometric redshifts. In particular, the sample in UDS now contains 257 H$\alpha$ emitters over 0.52 deg$^2$. The very small reduction in area simply results from the overlap with deep mid-infrared data used for deriving the new improved photometric redshifts -- this reduction places 3 sources from the initial sample outside the coverage. The remaining 6 sources excluded from the S09 sample were removed on the basis that the new photometric redshifts clearly place those sources at $z\approx2.2$ (2) and $z\approx1.45$ (4) (they are therefore identified as candidate [O{\sc ii}]\,3727 and [O{\sc iii}]\,5007/H$\beta$ emitters, respectively). For COSMOS, the sample of H$\alpha$ emitters is the same as in S09 (477 emitters over 0.76 deg$^2$), since the new photometric redshifts do not change any of the classifications (noting that spectroscopic data had already been used to produce a cleaner sample in S09).

As the samples were obtained in two of the best studied square degree areas, a wealth of multi-wavelength data are available. These include deep broad-band imaging from the ultra-violet (UV) to the infrared (IR) -- including deep $Spitzer$ data. These data make it possible to compute rest-frame luminosities for the sample of H$\alpha$ emitters at $z=0.84$. Here, rest-frame $B$ luminosities ($M_B$) are estimated by using $i^+$-band data (probing 4152.6 \AA\, at $z=0.84$) obtained with the {\sc subaru} telescope in both UDS and COSMOS. This is obtained by applying an aperture correction of -0.097 mag (to recover the total flux for magnitudes measured in 3 arcsec apertures) and a galactic extinction correction of -0.037 to $i^+$ magnitudes \citep{Capak}. It is assumed that all sources are at a luminosity distance of 5367\,Mpc ($z=0.845$); this yields a distance modulus of 43.649. Rest-frame K luminosities ($M_K$) are computed following \cite{Cirasuolo08} by assuming the same luminosity distance and interpolating using deep 3.6\,$\mu $m and 4.5\,$\mu $m $Spitzer$ data.

Furthermore, the COSMOS field has been imaged by {\sc acs/hst}\rm, and the sample of H$\alpha$ emitters has been morphologically classified both automatically, with {\sc zest} \rm  \citep{Scarlata}, and visually, as detailed in S09. The galaxies were classified into early-types, discs and irregulars and, independently (visually-only), into non-mergers, potential mergers and mergers (c.f. S09). The sample has also been investigated for AGN contamination both in S09, using emission-line diagnostics, and in \cite{Garn09} making use of a wide variety of techniques.

Figure \ref{on_sky} presents the distribution of the H$\alpha$ emitters at $z=0.84$ for the UDS and COSMOS fields. The panels visually indicate how these emitters cluster across these two regions of the sky relative to a simple photometric redshift selected population at the same redshift, selected with $0.82<z_{\rm photo}<0.87$.

\subsection{The random sample}\label{corr_funct}

Random catalogues are essential for robust clustering analysis and an over- or under-estimation of the clustering amplitude can easily be obtained if one fails to produce accurate random catalogues. These are produced by generating samples with 100 times more sources than the real data and by distributing those galaxies randomly over the geometry corresponding to the survey's field-of-view. This takes into account both the geometry of the fields and the removal of masked areas (due to bright stars/artefacts) -- see discussion of masked regions in S09 for more details.

While the survey is fairly homogeneous in depth, there are some small variations from chip to chip and field to field reaching a maximum of $\sim0.2$ mag (NB$_J$ of 21.5 to 21.7 mag). This corresponds to a maximum variation in luminosity [in $\log(L)$] of $\sim0.1$. The observed luminosity function presented in S09 shows that going deeper by 0.1 in $\log(L)$ increases the number count by $\sim20$ per cent. This can have an effect on estimating the clustering, although simulations indicate that this is smaller than the measured errors. The final random catalogues were created reproducing source densities varying in accordance with the measured depths and the S09 number counts.

\section{THE CLUSTERING PROPERTIES OF H$\alpha$ EMITTERS}

\subsection{The two-point $\bf \omega(\theta)$ correlation function at $\bf z=0.84$} \label{omega084}

In order to evaluate the two-point angular correlation function, the minimum variance estimator suggested by \cite{Landy1993} is used:

\begin{equation}
  \omega(\theta)= 1+ \left(\frac{N_R}{N_D}\right)^2 \frac{\rm DD(\theta)}{\rm RR(\theta)}-2\frac{N_R}{N_D}\frac{\rm DR(\theta)}{\rm RR(\theta)},
\end{equation}\label{ee}
where ${\rm DD(\theta)}$ is the number of pairs of real data galaxies within ($\theta, \theta+\delta\theta$), ${\rm DR(\theta)}$ is the number of data-random pairs and ${\rm RR(\theta)}$ is the number of random-random pairs. ${N_R}$ and ${N_D}$ are the number of random and data galaxies in the survey. Errors are computed using the Poisson estimate \citep{Landy1993}:

\begin{equation}
  \Delta\omega(\theta)= \frac{1+\omega(\theta)}{\sqrt{\rm DD(\theta)}}.
\end{equation}

\begin{figure}
\centering
\includegraphics[width=8.2cm]{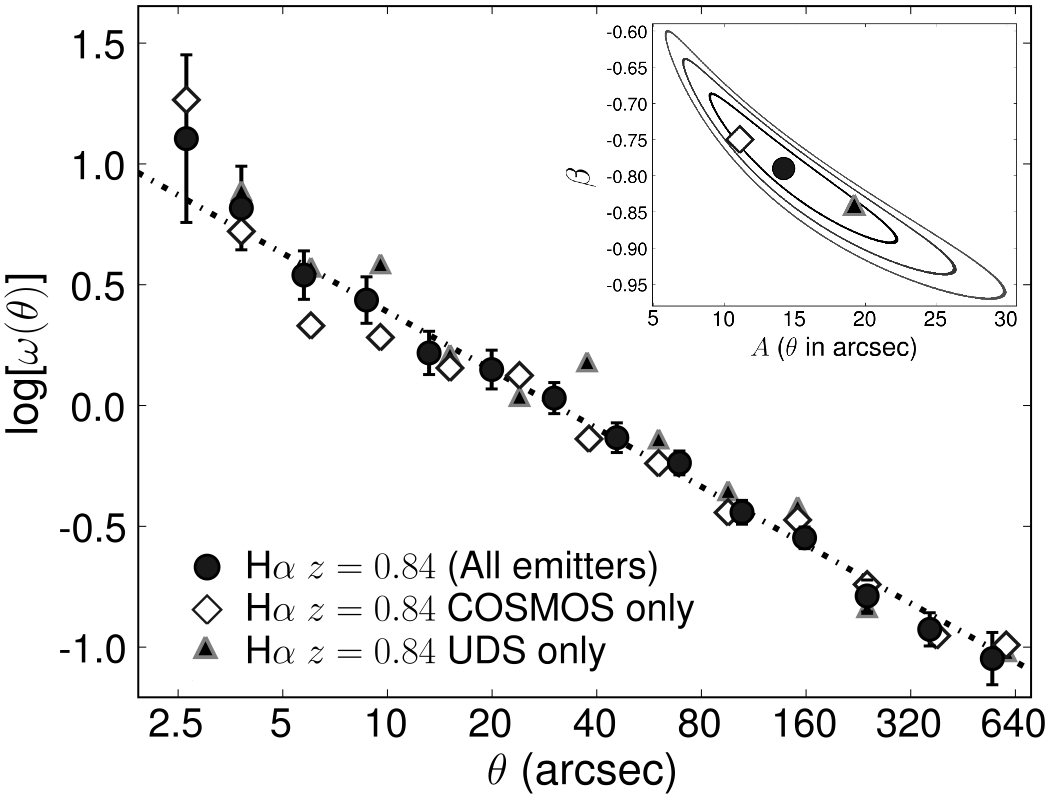}
\caption[wteta]{The two-point angular correlation function for the entire sample and for the COSMOS and UDS fields separately for angular separations up to 600 arcsec, and the 1$\sigma$, 2$\sigma$  and 3$\sigma$ co-variance contours when fitting both $A$ and $\beta$ for the entire sample compared to the results of each field. These results show a very good agreement between COSMOS and UDS, revealing that H$\alpha$ emitters in completely different regions of the sky cluster with comparable amplitudes and power-law slopes. The results also show that there is no clear departure from the power-law for this sample of H$\alpha$ emitters down to $5''$, implying that the 2-halo contribution to $\omega(\theta)$ is dominant for scales larger than $\sim$50\,kpc, but there is tentative evidence of a significant 1-halo contribution to $\omega(\theta)$ for smaller scales. \label{wtet_UDS_COSMOS}}
\end{figure}

The angular correlation function is computed for the entire sample of H$\alpha$ emitters\footnote{Due to the finite area probed, the measured clustering amplitude will be underestimated by an amount $C$ \citep[known as the integral constrain; c.f.][]{Roche} which depends on the assumed true power-law and the probed area. This is estimated as $C=0.0023$ for the entire survey area and it only represents $\approx0.01$ per cent change in $A$, but it is still included, since this becomes significant for obtaining $\omega(\theta)$ at large separations; c.f. Figure \ref{accurate_limb}.}, as well as for COSMOS and UDS separately, and it is found to be very well-fitted by a power-law with the form $A\theta^{\beta}$ with $\theta$ in arcsec. The power-law fit is obtained by determining $\omega(\theta)$ 2000 times using different random samples and a range of bin widths ($\Delta\log\theta=0.1-0.3$, randomly picked for each determination) and by performing a $\chi^2$ fit to each of these (over $5<\theta<600$ arcsec, corresponding to 38.2\,kpc to 4.5\,Mpc at $z=0.845$; these correspond to angular separations for which fitting $\omega(\theta)$ with one single power-law is appropriate; see details in Section \ref{accurate}). The results are shown in Figure \ref{wtet_UDS_COSMOS} and presented in Table \ref{clustering_wtheta}; they imply $A=14.1\pm3.9$ and $\beta=-0.79\pm0.06$ for the entire sample (the errors present the 1$\sigma$ deviation from the average). This reveals a very good agreement with the fiducial value, $\beta=-0.8$.

Recent studies \citep[e.g.][]{Ouchi_dep} have found a transition between small and large-scale clustering (corresponding to one- and two-halo clustering, respectively) at high redshift, manifested as a clear deviation from the power-law at small scales (smaller than $\approx50$\,kpc for Lyman-break galaxies at $z\sim4$, for example). While there is no definitive evidence for the one-halo term for the H$\alpha$ sample presented here (especially for separations larger than $\sim5''$ -- corresponding to 38\,kpc), for smaller separations the results point towards a departure from the power-law, but only at $\approx1\sigma$ level. Therefore, a larger sample is required to reliably determine $\omega(\theta)$ down to the smallest scales and constrain the contribution of the one-halo term.

The best power-law fit seen in Figure \ref{wtet_UDS_COSMOS} also reveals a good agreement both at small and larger scales between the COSMOS and UDS fields (see also Table 1); the UDS field presents a slightly higher clustering amplitude, but consistent within 1$\sigma$ (fixing $\beta=-0.8$). Note that the H$\alpha$ luminosity function derived in S09 for each field also revealed a good agreement between the two fields. These results are consistent with $\sim0.6-0.8$ deg$^2$ fields being sufficient to overcome most of the effects of cosmic variance when conducting clustering analysis with narrow-band surveys at $z\sim1$ -- although the agreement could be caused by chance.

\begin{table}
 \centering
  \caption{The power-law fit parameters which best describe $\omega(\theta)$ for H$\alpha$ emitters at $z=0.84$, resulting from computing $\omega(\theta)$ 2000 times. These were obtained over $5<\theta<600$ arcsec, corresponding to 38\,kpc to 4.5\,Mpc at $z=0.84$, avoiding both the possible one-halo clustering contribution (at the smallest scales) and the break of the Limber's approximation (at the largest scales). Note the degeneracy between $A$ and $\beta$ in Figure \ref{wtet_UDS_COSMOS}. $A_{\beta=-0.8}$ is obtained by fixing $\beta=-0.8$, the fiducial value and in excellent agreement with what has been found for the entire sample with or without possible/likely AGN contamination. An error of 20 per cent is added in quadrature to random error in $\Delta A_{\beta=-0.8}$ to account for cosmic variance for the entire sample (or 25 per cent when considering just one sub-field -- COSMOS or UDS, c.f. Section \ref{cosmic084}). }
  \begin{tabular}{@{}ccccc@{}}
  \hline
   \bf Sample & \bf Number & \bf A & \bf $\beta$ & \bf $A_{\beta=-0.8}$ \\
   \bf ($z=0.84$) & \# & ($\theta$ in arcsec) & & ($\theta$ in arcsec)  \\
  \hline
All Emitters & 734 & $14.1\pm3.9$ & $-0.79\pm0.06$ & $14.2\pm3.1$   \\
  \hline
No AGN & 660 & $13.6\pm4.4$ & $-0.78\pm0.07$ & $14.1\pm3.0$ \\
No likely AGN & 700 & $15.4\pm4.0$ & $-0.81\pm0.06$ & $14.6\pm3.2$ \\
  \hline
COSMOS & $477$ & $11.1\pm3.0$ & $-0.75\pm0.08$ & $13.8\pm3.7$ \\
UDS & $257$ & $19.2\pm8.9$ & $-0.84\pm0.15$ & $18.4\pm5.5$ \\
  \hline
\end{tabular}
\label{clustering_wtheta}
\end{table}

\subsubsection{Robust error estimation and the effect of cosmic variance} \label{cosmic084}

\begin{figure}
\centering
\includegraphics[width=8.2cm]{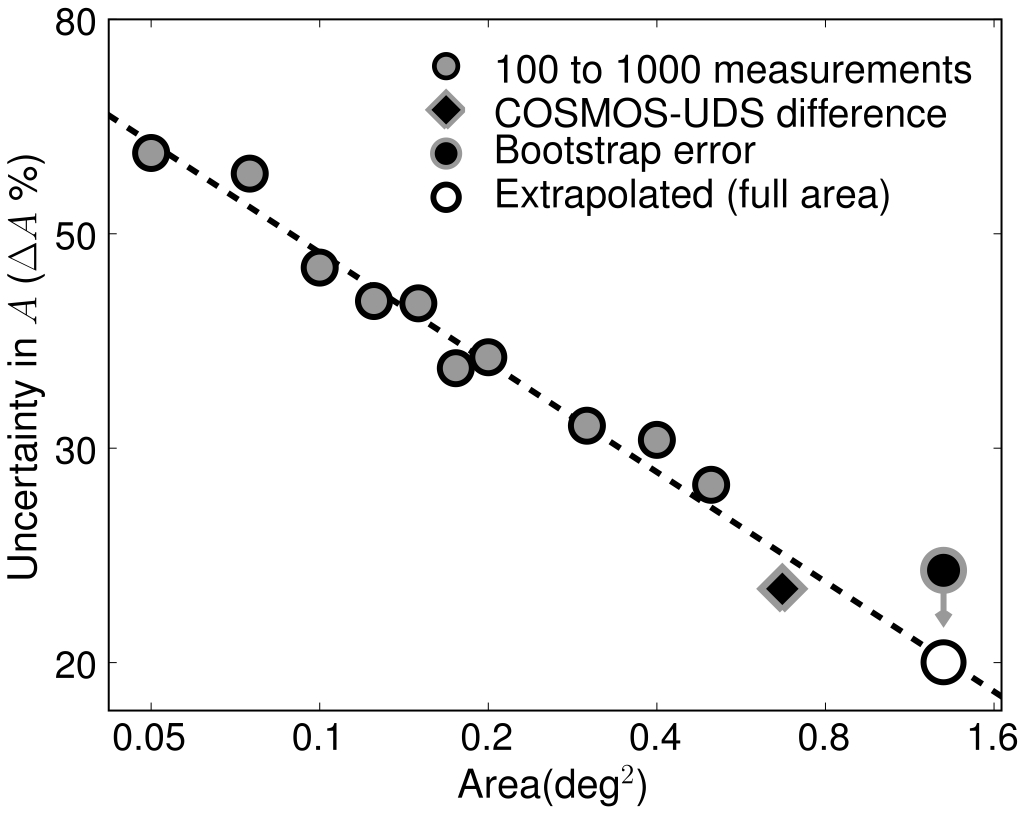}
\caption[wteta]{The uncertainty in $A$ (per cent; $\Delta A=\sigma/A\times100$, with $\beta=-0.8$) derived from 100-1000 $\omega(\theta)$ measurements on sub-samples spanning different areas within the complete 1.3 deg$^2$ (up to 0.5 deg$^2$) -- which is considered to be mostly due to cosmic variance. This reveals that $\Delta A$ decreases with increasing area and this trend is very well fitted by a simple power-law ($20\times\theta^{-0.35}$ per cent, with $\theta$ in deg$^2$, shown by the dashed line). The simple standard deviation between results from the COSMOS and UDS fields is also shown for comparison. The error estimated from the bootstrap analysis and that given by extrapolating the fitted power-law are also shown for the full area. \label{cosmic_var}}
\end{figure}

\begin{figure}
\centering
\includegraphics[width=8.2cm]{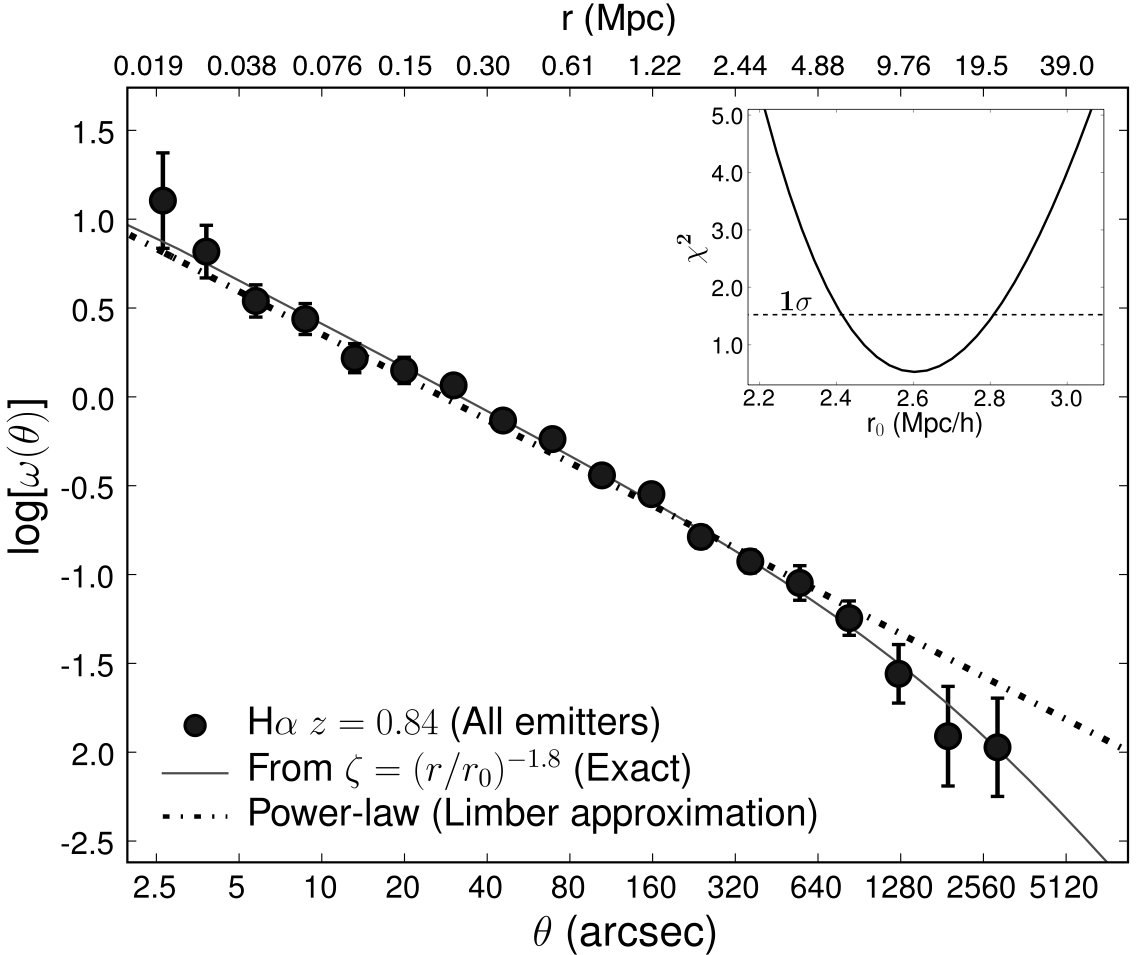}
\caption[wteta]{The two-point angular correlation function for the entire sample, compared with the simple power-law fit expected from the Limber's approximation. These are compared with the best (from a $\chi^2$ fit -- shown in the figure) exact angular correlation function obtained with the narrow-band filter profile for a spatial correlation function given by the power-law $\xi=(\rm r/r_0)^{\gamma}$, with $\gamma=-1.8$ and $\rm r_0=2.61\pm0.13$\,h$^{-1}$\,Mpc (not corrected for contamination; errors directly from $\chi^2$), revealing an excellent agreement with the data. This also reveals the regime for which the Limber's approximation breaks for this particular case ($\sim600$\,arcsec for a $\sim15$\,per cent difference; c.f. Simon 2007 for a general analysis of the typical separation at which Limber's approximation breaks down). \label{accurate_limb}}
\end{figure}

The survey covers 1.3 deg$^2$ in two completely independent fields; this is a fundamental advantage over smaller surveys which only probe one single field, as it allows, in principle, a reliable estimate of possible errors due to cosmic variance. In order to achieve this, $\omega(\theta)$ is estimated over randomly picked square areas of different sizes (from 0.05 deg$^2$, corresponding to the area probed by one WFCAM chip, up to 0.5 deg$^2$) in COSMOS and UDS. The minimum number of emitters ranges from 15 to 65 within the smaller areas used (0.05 deg$^2$). Either 100 and 1000 randomly chosen regions are considered for each area (100 for 0.3-0.5 deg$^2$ and 1000 for smaller areas) and a power law is fitted to each $\omega(\theta)$ determination by fixing $\beta=-0.8$. For each area, the standard deviation on the amplitude $A$ is used to quantify the uncertainty. The results are shown in Figure \ref{cosmic_var} and demonstrate that the measured standard deviation is effectively reduced with area. The error in $A$ (per cent) can be approximated by a power law of the form $20\times\theta^{-0.35}$ with $\theta$ in deg$^2$; extrapolating this suggests an error slightly lower than $\sim20$ per cent for the total area of the survey, and also agrees with the small difference found between COSMOS and UDS (see Figure \ref{cosmic_var}). Furthermore, this suggests that one requires areas of $\sim7$\,deg$^2$ (the HiZELS survey target) or more to reduce the effects of cosmic variance to less than 10 per cent.

These estimates could still slightly underestimate the errors, mostly because at larger areas the number of approximately independent areas is strongly reduced. An alternative estimate of the clustering uncertainty can be derived using a bootstrap analysis. Since this often leads to a slight over-estimation of the errors, it can be combined with the previous analysis to give an approximate range for the expected error. The analysis is done by naturally dividing the complete sample into 26 regions of $\sim$0.05 deg$^2$ each (these are the minimum individual probed areas as these are covered by one WFCAM chip) and computing $\omega(\theta)$ with 26 randomly picked regions each time for 10000 times. Fitting all realizations with a power law (fixing $\beta=-0.8$) results in a distribution with a standard deviation of 23 per cent in the clustering amplitude, $A$. When compared and combined with the previous estimation ($\approx18$ per cent error), it suggests an error of $\sim20$ per cent in the clustering amplitude; this will be added in quadrature to the directly calculated errors in $A$.

\subsubsection{AGN contamination}\label{morph_dep}

Some of the H$\alpha$ emitters are likely to contain an AGN. By using emission-line ratio diagnostics for a small subset of emitters in $z$COSMOS, S09 estimated a contamination of $\sim$15 per cent AGN in the sample. More recently, Garn et al. (2009) performed an extended search for AGN within the sample using several methods for identifying those sources, such as radio, X-rays and mid-infrared colours. This resulted in the identification of a maximum of 74 AGN (40 in COSMOS and 34 in UDS). From these, 34 are classified as likely AGN and the remaining 40 as possible AGN -- this corresponds to an estimated $\sim5$-11 per cent AGN contamination within the sample.

The AGN may well have different clustering properties from the star-forming population. However, the nature of these potential AGN contaminants is very unclear, particularly the origin of the detected H$\alpha$ emission. For example, they are mostly morphologically classified as irregulars and mergers and they span the entire luminosity range; it thus seems that although some of them might have their H$\alpha$ emission powered by the AGN, they may well be undergoing significant star-formation. Indeed, recently, \cite{shi} presented a study of unambiguous AGN at $z\sim1$, showing that at least half of the sample shows clear signatures of intense star-formation.

In order to understand how AGN contaminants might affect the clustering measurements, the angular correlation function was calculated after removing all the potential AGN contaminants. This results in no statistical change in either $A$ or $\beta$ for the entire sample (see Table 1). Removing only the likely AGN leads to an identical result. However, the true AGN contaminants (and the possible bias they may introduce) might well have a larger effect on the results when studying the clustering as a function of host galaxy properties (presented in Section 4), since for certain host galaxy properties the AGN contamination may be significantly higher than in the sample as a whole.

To ensure that the analysis is really tackling the AGN contamination and to robustly determine the clustering properties of each sample being studied (and test trends), all the clustering properties in this paper (for the $z=0.84$ sample and sub-samples) are derived from a combination of measurements from samples with and without possible AGN contaminants. In practice, $\omega(\theta)$ is computed 500 times each using i) all the emitters in the sample, ii) removing likely AGN, and iii) removing all (likely plus possible) AGN. $\chi^2$ fits are obtained for each realization of each sample and the total resulting distribution is used for the analysis. This also allows to carefully confirm that there is a good degree of consistency between the 3 different distributions and in no case is the difference between samples with and without AGN larger than the 1$\sigma$ errors. This mostly results in estimating larger errors than those which would be obtained from either simply excluding all AGN or not dealing with AGN contamination, and therefore represents a conservative approach.

\subsection{Real-space correlation} \label{Reals}

\subsubsection{First order approximation: Limber's equation} \label{limber}

The real-space correlation, r$_0$, is a very useful description of the physical clustering of galaxies when the spatial correlation function is well-described by $\xi=(\rm r/\rm r_0)^{\gamma}$. The inverse Limber transformation \citep{Peebles} can be easily used to obtain an approximation of the spatial correlation function\footnote{Limber's equation is an approximation that breaks down for large angular separations when one uses a narrow filter, but can still be used to obtain at least a first approximation of r$_0$ within $\approx15$ per cent for the NB$_J$ filter used; see Section 3.2.2 for details.} from the angular correlation function, provided the redshift distribution is known. For narrow-band surveys, the expected redshift distribution depends solely on the shape of the narrow-band filter profile; this can be well approximated by a Gaussian\footnote{A simpler way to model the narrow-band filter is to assume it is a top-hat; this results in a H$\alpha$ redshift distribution of $0.845\pm0.015$. The two approaches produce results for different r$_0$ determinations which are consistent within 5 per cent.}, which, for the H$\alpha$ emission-line, corresponds to a redshift distribution centered at $z=0.845$ with $\sigma=0.0075$. This can be compared with the redshift distribution from H$\alpha$ emitters confirmed by $z$COSMOS (93 sources) which has a mean redshift of 0.844 and $\sigma=0.0076$; the excellent agreement reveals that the real redshift distribution should be very close to the one assumed.

In order to calculate the real space correlation length, r$_0$, and when performing the de-projection analysis, it is assumed that the redshifts are drawn from this Gaussian distribution and that the real-space correlation function is independent of redshift \citep[c.f. G08;][]{Kovac07} over the redshift range probed; this is a very good approximation, since the redshift distribution is extremely narrow. Contamination by sources which are not H$\alpha$ emitters at $z=0.84$ could have a significant effect on r$_0$, since the real redshift distribution will be different from the one assumed. Nevertheless, in S09 the sample was studied to find that only 2 out of 90 sources with spectroscopic redshifts that had been photometrically selected as H$\alpha$ emitters were not real H$\alpha$ emitters (one [S{\sc ii}]6717 emitter at $z=0.79$ and a [O{\sc iii}]\,5007 emitter at $z=1.42$, which were then removed from the sample). This conclusion is drawn from the analysis of limited spectroscopic data ($\sim20$ per cent of the HiZELS sample in COSMOS), but contamination at that level will only lead to underestimating r$_0$ by a maximum of 6 per cent\footnote{Assuming that contaminants will not cluster significantly, the contamination fraction of $f$ will result in a maximum underestimation of $A$ given by (1-$f$)$^2$, and an underestimation of r$_0$ given by $\sim$\,(1-$f$)$^{2/|\gamma|}$.}. Throughout this paper, a 5 per cent correction is applied when computing r$_0$ to account for the expected contamination.

Computing r$_0$ for each realization of $\omega(\theta)$ fitted with a power-law with $\beta=-0.8$ results in r$_0=2.5\pm0.3$\,h$^{-1}$\,Mpc for the entire sample, or r$_0=2.6\pm0.3$\,h$^{-1}$\,Mpc when accounting for 5 per cent contamination. Note that the 20 per cent error in $A$ due to cosmic variance results in an error of 11 per cent in r$_0$, and this is added in quadrature. However, as mentioned before, Limber's equation is only an approximation which can potentially result in significant errors for narrow-band surveys at high redshift. Section \ref{accurate} describes how r$_0$ is robustly calculated by fully de-projecting $\omega(\theta)$.

\subsubsection{Accurate determination of r$_0$ for narrow-band surveys} \label{accurate}

The errors introduced by the Limber's approximation are tackled by numerically integrating the exact equation connecting the spatial and angular correlation functions \citep[following][]{Simon07}. This relation implies that spatial correlation functions described by  $\xi=(\rm r/r_0)^{\gamma}$ are projected as angular correlation functions with slopes $\beta=\gamma+1$ for small scales and $\beta=\gamma$ for large angular separations when using a narrow filter -- see full discussion in Simon (2007).

Here, it is assumed that the spatial correlation function is given by the power-law $\xi=(\rm r/r_0)^{\gamma}$ (as this is able to reproduce the observed $\omega(\theta)$ very well with $\gamma=-1.8$), and that the narrow-band filter profile is described by a Gaussian as detailed in Section \ref{limber}. The following is then numerically integrated for the same angular separations as for the data and a $\chi^2$ fit is done for $\omega(\theta)$ around the value of r$_0$ obtained in the previous Section:

\begin{equation}
  \Delta\omega(\theta)= \psi^{-1} \int_{0}^{+\infty} \int_{s\sqrt{2\phi}}^{2s}\frac{2{\rm f_S}(s-\Delta){\rm f_S}(s+\Delta)}{ {\rm R}^{-\gamma-1}r_0^{\gamma} \Delta }\mathrm{d}{\rm R}\mathrm{d}s,
\end{equation}
where $\psi=1+\cos\theta$, $\phi=1-\cos\theta$, $\Delta=\sqrt{({\rm R}^2-2s^2\phi)/(2\psi)}$ and ${\rm f_S}$ is the profile of the filter being used in co-moving distance (assumed to be a Gaussian distributed value with an average of 2036.3\,h$^{-1}$\,Mpc and $\sigma=14.0$\,h$^{-1}$\,Mpc). The $\chi^2$ fitting results in a much more robust estimate of r$_0$ and allows the use of $\omega(\theta)$ up to larger angular separations than 600 arcsec, a regime for which a single power-law starts to become inadequate (as $\beta$ changes from $\gamma+1$ to $\gamma$, c.f. Figure 4). For the entire sample, however, this leads to little change: $\rm r_0=2.7\pm0.3$\,h$^{-1}$\,Mpc (this includes a 5 per cent correction for contamination, while the errors also include cosmic variance -- this error estimation has been discussed in Section \ref{cosmic084}). Indeed, whilst Limber's equation breaks down for large galaxy separations, it is shown to do well as long as only smaller angular distances are considered. Note that real-space correlations for different samples (see Table \ref{clustering_table}) are computed in the following sections using the same procedure described here.

\subsubsection{The dependence of the redshift distribution on limiting line luminosity for narrow-band surveys} \label{Reals}

Since the narrow-band filter profile is not a perfect top-hat, emitters with different H$\alpha$ luminosities are detected over a slightly different range of redshifts. As the exact relation between $\xi(\rm r)$ and $\omega(\theta)$ depends on the redshift distribution (given by the filter profile), assuming the same redshift distribution for different luminosity limits can lead to biases, especially when looking at a possible relationship between the clustering amplitude and H$\alpha$ luminosity.

This potential bias is studied by performing the same simulations described in S09. Briefly, the H$\alpha$ luminosity function presented in S09 is used to generate a fake population of emitters equally distributed over a wider range of redshifts than those probed by the narrow-band filter and this is used to look at the recovered redshift distribution of sources as a function of measured luminosity. The redshift distribution is found to become continuously narrower with increasing H$\alpha$ luminosity limit\footnote{Although intrinsically more luminous sources are detectable over a wider redshift range, their detection in the wings of the filter leads to an under-estimation of their luminosity; galaxies are only measured to be luminous in H$\alpha$ when the emission line is being detected near the peak of the filter profile.}, although all the distributions can be equally well-fitted by a Gaussian. The variation can be written as a function of observed H$\alpha$ flux as:
\begin{equation}
   \sigma=-\eta\times(\log F_{ \rm {H\alpha}(limit)}-\log F_0)+\sigma_0,
\end{equation} \label{eq2}
where $F_0$ is a flux limit at which the redshift distribution is relatively well understood and $\sigma_0$ is the width of  the redshift distribution at that flux limit; for the NB$_J$ filter, $\eta=0.00117$, as derived from simulations.

Over the luminosity range of the sample (and for the assumed luminosity function and the real filter profile), neglecting this effect results in over-estimating r$_0$ by a maximum of 8 per cent; this is only found when computing r$_0$ for a sample containing the brightest 5 per cent emitters (the $\approx$ 40 brightest emitters, for which $\sigma\approx0.0065$). Nevertheless, narrow-band surveys which span a much wider luminosity range will be much more sensitive to this and line-luminosity trends could naturally arise from this bias.

This luminosity dependence is fully taken into account when computing r$_0$ for sub-samples defined by different H$\alpha$ luminosity/SFR limits. For the analysis of other sub-samples, if the H$\alpha$ luminosity distribution does not present any clear off-set from that of the entire survey, it is assumed that the redshift distribution of those sub-samples is very well approximated by the complete redshift distribution.

\section{The clustering dependences on host galaxy properties}\label{lum_dep}

As discussed in the Introduction, in the local Universe the clustering has been shown to depend on several host galaxy properties. At $z=0.24$, \cite{Shioya2008} have shown that the clustering of H$\alpha$ emitters seems to be stronger for the most luminous sources in H$\alpha$, but so far testing and quantifying that clustering dependence at higher redshifts for H$\alpha$ emitters has not been possible, as surveys have lacked area and sample size. The HiZELS sample is large enough to achieve this.

\begin{figure}
\centering
\includegraphics[width=8.2cm]{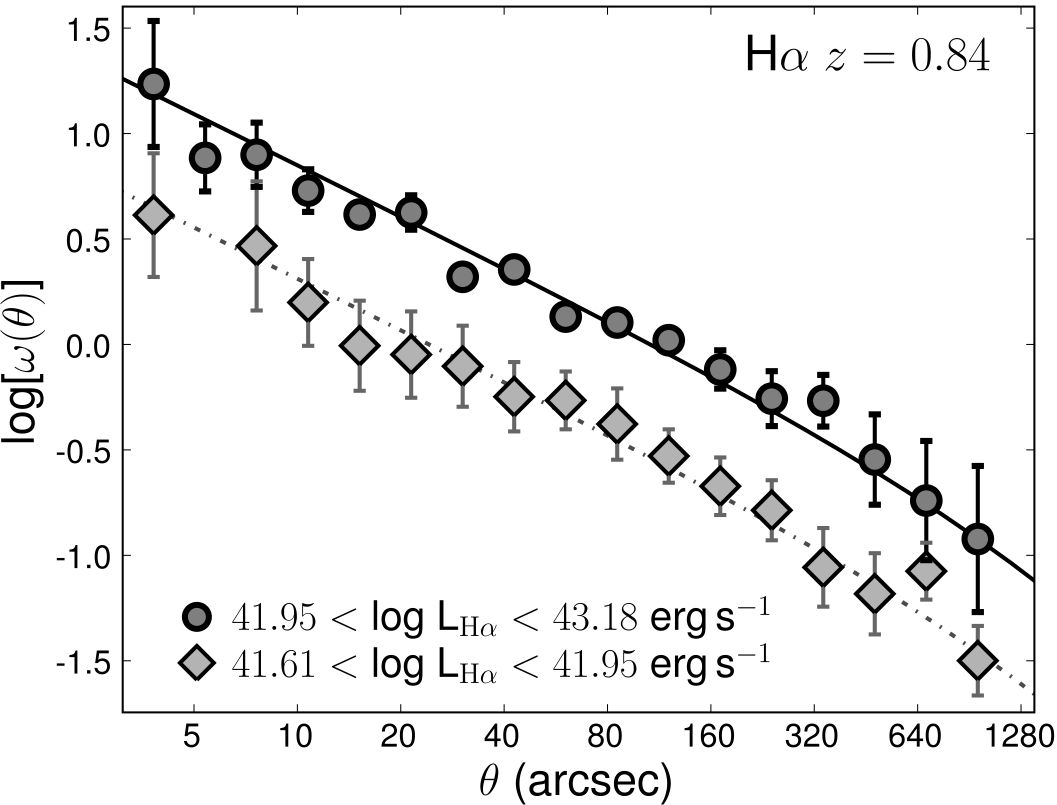}
\caption[wteta_subsam]{The two-point angular correlation function for sub-samples obtained by splitting the sample in 2 halves based on H$\alpha$ luminosity/SFR. These clearly show that the clustering amplitude is higher for galaxies with higher H$\alpha$ luminosities, implying that the most active star-forming galaxies are more clustered than the least active. Lines show the best $\chi^2$ to the exact relation between $\xi(r)$ and $\omega(\theta)$ by fixing $\gamma=-1.8$; which (after applying all corrections already detailed) result in r$_0=4.84\pm0.58$\,$\rm h^{-1}$\,Mpc for the brightest emitters and $2.52\pm0.41$\,$\rm h^{-1}$\,Mpc for the faintest half (in H$\alpha$ luminosity) of the sample. \label{wtet_sub_sam}}
\end{figure}

\begin{figure}
\centering
\includegraphics[width=8.2cm]{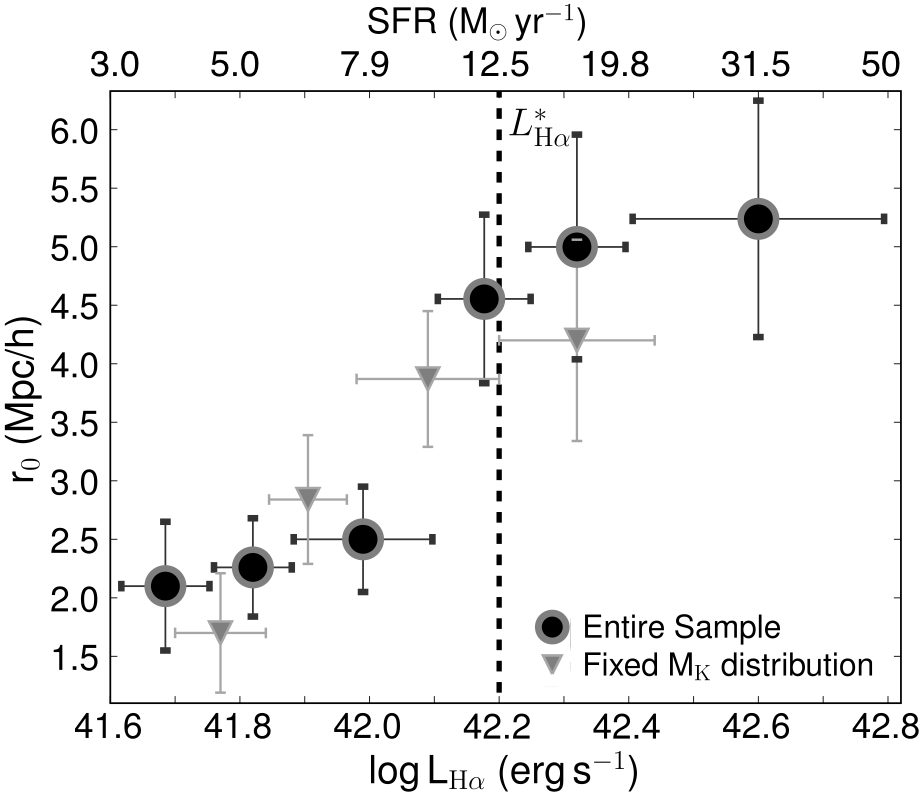}
\caption[wteta]{The dependence of the clustering length, r$_0$, with H$\alpha$ luminosity at $z=0.84$. This clearly reveals that galaxies presenting higher H$\alpha$ luminosities/SFRs are more clustered, and that galaxies above the break in the H$\alpha$ luminosity function ($L_{\rm H\alpha}^*$) are much more clustered than those below. This dependence is not a result of the correlation between SFR (H$\alpha$ luminosity) and stellar mass ($M_K$, see Figure \ref{matchMK_Ha}), as r$_0$ correlates as well with H$\alpha$ luminosity for sub-samples with the same $M_K$. \label{r_0_Lalpha}}
\end{figure}

\begin{figure*}
\centering
\includegraphics[width=17.7cm]{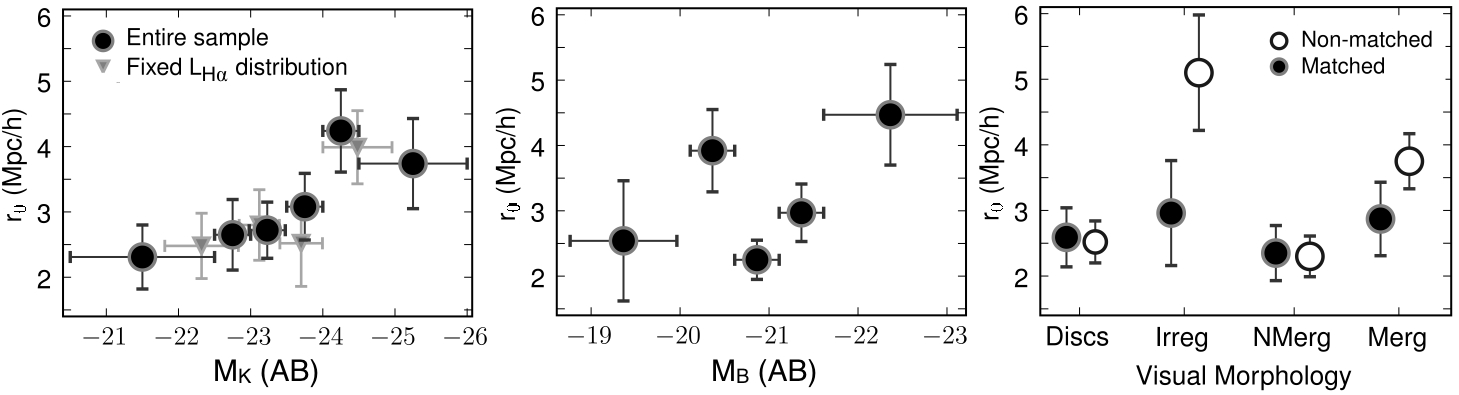}
\caption[wteta]{The real-space correlation length as a function of rest-frame $K$ luminosity (left panel), rest-frame $B$ luminosity (middle panel) and visual morphology (right panel). These results show that star-forming galaxies cluster more with increasing rest-frame $K$  luminosity (which traces stellar mass), with this trend being maintained (but slightly weakened) even when one uses sub-samples that have the same H$\alpha$ luminosity distribution. No statistically significant correlation between r$_0$ with rest-frame $B$ luminosity is found. Morphologically classified irregulars and mergers cluster more strongly than discs and non-mergers, but this is due to a different H$\alpha$ luminosity and $M_K$ distribution within each population; matching these distributions results in a similar r$_0$. \label{r_0_dep}}
\end{figure*}

The complete sample is divided into several sub-samples in order to investigate the clustering as a function of fundamental observable host galaxy properties. These include i) H$\alpha$ luminosity corrected for extinction as in S09 (1 mag), ii) rest-frame $K$ luminosity ($M_K$), iii) rest-frame $B$ luminosity ($M_B$) and iv) morphological class (discs, irregulars; non-mergers, mergers). The $\omega(\theta)$ correlation function is computed for each sub-sample (a single example is shown in Figure \ref{wtet_sub_sam}, presenting $\omega(\theta)$ for the brightest and faintest halves of the sample with respect to H$\alpha$ luminosity). The approach detailed in the previous sections is then used by firstly computing r$_0$ using Limber's approximation power-law fit to $\omega(\theta)$ (fixing $\beta=-0.8$ and restricting the analysis to $5<\theta<600$ arcsec), and then using the exact relation between the spatial and angular correlation functions to improve the r$_0$ estimation. This analysis also accounts for the potential AGN contamination in the sample, following the procedures described at the end of Section 3.1.2 (computing $\omega(\theta)$ with and without the possible AGN contaminants and using the combined distribution). The results obtained when splitting the sample into different sub-samples are presented in Table \ref{clustering_table}, Figure \ref{r_0_Lalpha} and Figure \ref{r_0_dep}. For sub-samples obtained from the complete survey area, an error of 11 per cent in r$_0$ is added in quadrature to the 1 $\sigma$ errors (see Section 3.1.1), while for sub-samples derived based on morphology a 14 per cent error is added in quadrature to $\Delta \rm r_0$ (for the COSMOS field only).

\begin{table}
 \centering
  \caption{The correlation length (fixing $\gamma=-1.8$ and corrected by 5 per cent for contamination) for H$\alpha$ emitters at $z=0.84$. Samples with fixed $M_K$ have $-24.44<M_K<-22.68$\,(AB) and those with fixed log\,L$_{\rm H\alpha}$ have $41.76<\rm log\,L_{\rm H\alpha}<42.22$ (erg\,s$^{-1}$). H$\alpha$ luminosities are given in erg\,s$^{-1}$, $M_K$ and $M_B$ are AB rest-frame luminosities. Unclassified sources or with no data available were not considered. Errors include cosmic variance (11 per cent of r$_0$ for the entire sample and 14 per cent of r$_0$ for samples of each individual field).}
  \begin{tabular}{@{}ccc@{}}
  \hline
   \bf Sub-sample & \bf Number & \bf r$_{0}$ \\
   \bf (H$\alpha$ $z=0.84$) & \# & ($\rm h^{-1}$\,Mpc)  \\
  \hline
\bf All Emitters & 734 & $2.74\pm0.29$   \\
Pure Star-forming & 660 & $2.61\pm0.28$   \\
COSMOS & $477$ & $2.67\pm0.37$ \\
UDS & $257$ & $3.13\pm0.52$ \\
  \hline
$41.95<$ log L$_{\rm H\alpha}$ $<43.18$ & 367 & $4.84\pm0.58$  \\
$41.61<$ log L$_{\rm H\alpha}$ $<41.95$ & 367 & $2.52\pm0.41$ \\
  \hline
$42.40<\rm log\,L_{\rm H\alpha}<42.80$ & $46$ & $5.24\pm1.01$  \\
$42.25<\rm log\,L_{\rm H\alpha}<42.40$ & $55$ & $5.03\pm0.96$   \\
$42.10<\rm log\,L_{\rm H\alpha}<42.25$ & $92$ & $4.55\pm0.72$   \\
$41.88<\rm log\,L_{\rm H\alpha}<42.10$ & $243$ & $2.50\pm0.45$    \\
$41.75<\rm log\,L_{\rm H\alpha}<41.88$ & $173$ & $2.26\pm0.42$    \\
$41.62<\rm log\,L_{\rm H\alpha}<41.75$ & $125$ & $2.13\pm0.55$    \\
$42.20<\rm log\,L_{\rm H\alpha}<42.44$ (fixed  $M_K$) & 56 & $4.20\pm0.86$   \\
$41.98<\rm log\,L_{\rm H\alpha}<42.20$ (fixed  $M_K$) & 117 & $3.87\pm0.58$   \\
$41.83<\rm log\,L_{\rm H\alpha}<41.98$ (fixed  $M_K$) & 125 & $2.84\pm0.55$   \\
$41.71<\rm log\,L_{\rm H\alpha}<41.83$ (fixed  $M_K$) & 117 & $1.70\pm0.51$   \\
  \hline
$-26.1<M_K<-24.5$ & 82 & $3.74\pm0.69$  \\
$-24.5<M_K<-24.0$ & 116 & $4.24\pm0.63$  \\
$-24.0<M_K<-23.5$ & 145 & $3.08\pm0.51$   \\
$-23.5<M_K<-23.0$ & 145 & $2.72\pm0.43$   \\
$-23.0<M_K<-22.5$ & 117 & $2.65\pm0.54$   \\
$-22.5<M_K<-20.5$ & 116 & $2.31\pm0.49$   \\
$-24.96<M_K<-24.00$ (fixed  log\,L$_{\rm H\alpha}$) & 116 & $3.99\pm0.56$   \\
$-24.00<M_K<-23.40$ (fixed  log\,L$_{\rm H\alpha}$) & 108 & $2.52\pm0.62$   \\
$-23.40<M_K<-22.84$ (fixed  log\,L$_{\rm H\alpha}$) & 112 & $2.81\pm0.54$   \\
$-22.84<M_K<-21.81$ (fixed  log\,L$_{\rm H\alpha}$) & 111 & $2.48\pm0.50$   \\
  \hline
$-23.23<M_B<-21.61$ & $96$ & $4.47\pm0.77$  \\
$-21.61<M_B<-21.11$ & $186$ & $2.97\pm0.44$   \\
$-21.11<M_B<-20.61$ & $246$ & $2.25\pm0.30$    \\
$-20.61<M_B<-20.11$ & $145$ & $3.92\pm0.63$    \\
$-20.11<M_B<-18.71$ & $59$ & $2.54\pm0.92$    \\
  \hline
Discs (COSMOS) & $363$ & $2.52\pm0.32$  \\
Irregulars (COSMOS) & $68$ & $5.12\pm0.88$   \\
Discs (SFR \& $M_{K}$ match) & 109 & $2.59\pm0.45$  \\
Irregulars (SFR \& $M_{K}$ match) & 46 & $2.96\pm0.80$  \\
  \hline
Non-mergers (COSMOS) & $298$ & $2.30\pm0.31$  \\
Mergers (COSMOS) & $111$ & $3.75\pm0.50$    \\
Non-mergers (SFR \& $M_{K}$ match) & $128$ & $2.35\pm0.42$  \\
Mergers (SFR \& $M_{K}$ match) & $100$ & $2.87\pm0.56$  \\
  \hline
Bulge-dominated discs (COSMOS) & $97$ & $2.72\pm0.54$  \\
Disc-dominated discs (COSMOS) & $204$ & $2.51\pm0.39$   \\
  \hline
\end{tabular}
\label{clustering_table}
\end{table}

\subsection{H$\alpha$ luminosity/ star-formation rate}\label{SF_dep}

Following previous studies, such as \cite{Shioya2008}, one can do a simple splitting of the sample in two halves with the same number of emitters (367 with $41.95<$ log L$_{\rm H\alpha}$ $<43.18$ erg\,s$^{-1}$ and equal number with $41.61<$ log L$_{\rm H\alpha}$ $<41.95$ erg\,s$^{-1}$), and study the clustering properties of each of those samples. Figure \ref{wtet_sub_sam} presents the angular correlation function obtained for the brightest and faintest halves of the sample, revealing that the brightest H$\alpha$ emitters are much more clustered than the faintest ones. This is also very clear when one compares the spatial correlations obtained for each sample: while the brightest emitters in H$\alpha$ present r$_0=4.8\pm0.6$\,$\rm h^{-1}$\,Mpc, for the faintest emitters one finds $2.5\pm0.4$\,$\rm h^{-1}$\,Mpc. It should also be noted that because $\gamma$ is fixed, the significant difference in r$_0$ for the two samples is not a result of the degeneracy between $\gamma$ and r$_0$ (see Figure \ref{wtet_UDS_COSMOS} for a similar degeneracy between $\beta$ and $A$), contrarily to studies such as \cite{Shioya2008}, which allow for both $\gamma$ and r$_0$ to vary; such note is valid throughout this paper.

With the large sample of H$\alpha$ emitters obtained by HiZELS at $z=0.84$, it is possible to perform a much more detailed investigation into the relation between r$_0$ and H$\alpha$ luminosity by separating the emitters into a larger number of sub-samples. Figure \ref{r_0_Lalpha} and Table 2 show that r$_0$ increases by a factor of almost 3 from the faintest H$\alpha$ galaxies to the most active star-forming galaxies found above the break in the H$\alpha$ luminosity function, $L_{\rm H\alpha}^*$ ($10^{42.2}$\,erg\,s$^{-1}$, S09).

Whilst the increase of r$_0$ with L$_{\rm H\alpha}$ can be reasonably well-described by a straight-line fit (\,$\chi^2$ = 1.4\,), there are hints that there might be a stronger increase in r$_0$ around $L_{\rm H\alpha}^*$. This provides a description which is in line with previous studies, but gives an unprecedented degree of detail of the relation between r$_0$ and H$\alpha$ luminosity/star-formation rate. Moreover, the robustness of these results is also increased by including the small correction for the fact that samples with different luminosity limits will present a different redshift distribution just from considering the filter profile.

Recently, \cite{Orsi} made predictions for the clustering properties of H$\alpha$ emitters using two different versions of the {\sc galform} galaxy formation model. At $z=0.84$ and for the flux limit of S09, \cite{Orsi} predicts r$_0\approx4-5$\,h$^{-1}$\,Mpc. This shows that the disagreement between the data and the models found at $z=0.24$ and $z=0.4$ in \cite{Orsi} is also found at $z=0.84$. The models predictions would be consistent with the data for flux limits 2 times higher due to the strong clustering dependence with H$\alpha$ luminosity at $z=0.84$; however, none of the models is able to fully reproduce this r$_0$-L$_{\rm H\alpha}$ relation at the moment.

\subsection{Rest-frame continuum luminosity}\label{K_dep}

H$\alpha$ emitters presenting the highest rest-frame $K$ band luminosities ($M_K<-24$) are strongly clustered, and a general trend of an increasing r$_0$ with $K$ band luminosity is found (see left panel of Figure \ref{r_0_dep}), similarly to what has been found for other populations of galaxies at different redshifts. This suggests that this correlation must be valid regardless of the population being observed -- at least for the range of luminosities probed -- and seems to simply imply that galaxies with the highest stellar masses \citep[as these are expected to correlate very closely with $K$ luminosity, although the contribution from the thermally pulsing asymptotic giant branch (TP-AGB) phase of stellar evolution can lead to some scatter in the $M_K$ versus stellar mass relation; c.f.][]{Maraston} reside in the most massive haloes. However, it is also clear that the increase of r$_0$ with rest-frame $K$ luminosity, whilst continuous, is not as pronounced as the increase with star-formation rate.

In the local Universe, galaxies with higher $B$ band luminosities ($M_B$) are more clustered than galaxies with lower $M_B$. However, studying the clustering properties of H$\alpha$ emitters at $z=0.845$ as a function of rest-frame $B$ luminosity reveals no statistically significant trend ($<1\sigma$) (see middle panel of Figure \ref{r_0_dep}). These results are in reasonable agreement with recent studies \citep[e.g.][]{zCOSMOS_clustering}, which did not find any significant dependence of galaxy clustering on $B$ luminosity for a similar redshift range.

\subsection{Morphological class}\label{Morph_dep}

By splitting the sample into morphological classes (only for the COSMOS field [c.f. S09]), one finds that galaxies classified as irregulars are more clustered than those classified as discs; mergers have a measured r$_0$ which lies between these populations, but significantly above the non-mergers (see Table \ref{clustering_table} and right panel of Figure \ref{r_0_dep} -- open circles). However, S09 found that irregulars and mergers are typically brighter in H$\alpha$ and $M_K$ than discs (which completely dominate the faint-end of the H$\alpha$ luminosity function at $z\sim1$). Therefore, in order to understand if the clustering really does depend on the morphological class or if this is simply a secondary effect driven by star-formation rate and stellar mass dependencies, one needs to compare samples which are matched in log\,L$_{\rm H\alpha}$ and $M_K$. This is done by using the distribution of H$\alpha$ emitters in the $\log\,\rm L_{\rm H\alpha}$-$M_K$ plane and matching irregulars with discs and mergers with non-mergers. A $\Delta \log\,\rm L_{\rm H\alpha}<0.02$ \& $\Delta M_K<0.02$ criteria is used: these values were chosen to ensure that the samples are very well-matched in both H$\alpha$ luminosity and $M_K$ while still retaining the sample sizes required for the analysis. The population match used still results in a severe reduction of the larger disc and non-merger samples (along with a smaller reduction of the number of irregulars and mergers), but by matching these on the basis of $M_K$ and $\log\,\rm L_{\rm H\alpha}$ distribution, it is then possible to directly compare these populations on the basis of the morphological class only.

There is no significant difference in r$_0$ for the matched samples of irregulars and discs (see Figure \ref{r_0_dep} -- filled circles) within 1$\sigma$. The r$_0$ difference between mergers and non-mergers is also greatly reduced, although mergers are still found to be slightly more clustered than non-mergers at $\sim$ 1\,$\sigma$ level.

The sample of H$\alpha$ emitters with disc morphologies has also been classified according to how much disc/bulge dominated each galaxy is \citep[with {\sc zest; \rm c.f.}][]{Scarlata}. By computing the correlation length for disc dominated discs and bulge dominated discs, one finds that they present reasonably the same r$_0$ ($2.7\pm0.5\,\rm h^{-1}$\,Mpc for disc dominated and $2.5\pm0.4\,\rm h^{-1}$\,Mpc for bulge dominated galaxies). Therefore, the bulge-disk ratio of the studied star-forming disc galaxies does not have any significant effect on how these galaxies cluster at $z\sim1$.

These results show that H$\alpha$ luminosity and $M_K$ (probing stellar mass) are the key host galaxy properties driving the clustering of star-forming galaxies at $z\sim1$ and morphology is unimportant for the clustering of galaxies at high redshift, just as has been found in the local Universe \citep[e.g.][]{Skibba}.

\subsection{H$\alpha$ luminosity versus rest-frame $K$ luminosity}\label{SF_dep}

\begin{figure}
\centering
\includegraphics[width=8.2cm]{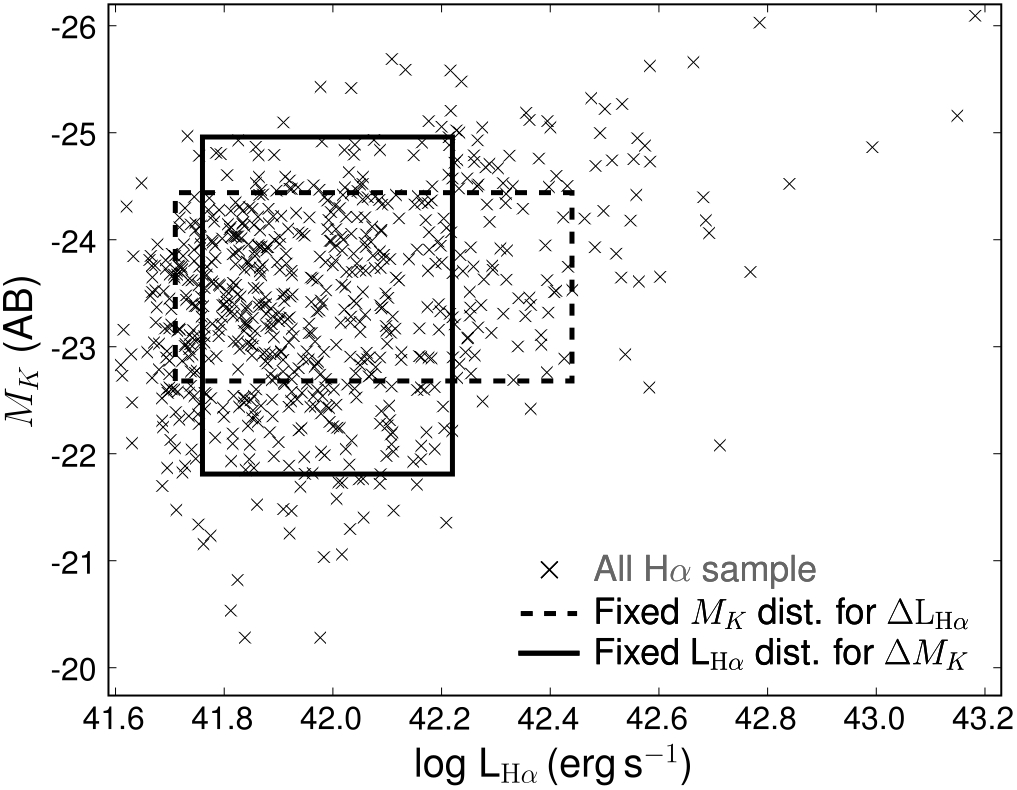}
\caption[wteta_subsam]{The complete 2-D $M_K$-log\,L$_{\rm H\alpha}$ space, showing the correlation between rest-frame $K$ luminosity and H$\alpha$ luminosity, compared to the regions used to select sub-samples with the same $M_K$ or L$_{\rm H\alpha}$ distributions. This clearly shows that without using sub-regions of this 2-D space, samples defined only on either L$_{\rm H\alpha}$ or $M_K$ will present different $M_K$ and L$_{\rm H\alpha}$ distributions, respectively, since these quantities are correlated. \label{matchMK_Ha}}
\end{figure}

It has been shown that r$_0$ is very well correlated with both H$\alpha$ luminosity (or star-formation rate) and rest-frame $K$ luminosity (tracing stellar mass). On the other hand, star-forming galaxies both in the local Universe and at high redshift reveal a correlation between both, with star-formation rate typically being higher for galaxies with higher stellar masses (see Figure \ref{matchMK_Ha}). Thus, how much is the correlation between r$_0$ and H$\alpha$ luminosity driven by different rest-frame $K$ luminosity distributions, and vice-versa?

In order to clearly test if both relations hold -- or whether they are a result of a single strong relation between r$_0$ and either H$\alpha$ luminosity of rest-frame $K$ luminosity -- new sub-samples are derived. To do this, two sub-regions are defined in the 2-D L$_{\rm H\alpha}$--$M_K$ space (see Figure \ref{matchMK_Ha}). To test the r$_0$-L$_{\rm H\alpha}$ correlation at a fixed $M_K$, the region $-24.44<M_K<-22.68$\,(AB) and $41.71<\rm log\,L_{\rm H\alpha}<42.44$ erg\,s$^{-1}$ (corresponding to $4.1<SFR<21.8$ \,M$_{\odot}$\,yr$^{-1}$) is considered (see Figure \ref{matchMK_Ha}). This region contains 409 H$\alpha$ emitters, which can be divided into 4 sub-samples on the basis of H$\alpha$ luminosities with the same distribution of $M_K$ (means of $-23.6$, $-23.7$, $-23.6$ and $-23.7$ with standard deviations of 0.5 in all instances; ordered from highest to lowest in respect to H$\alpha$ luminosity). Similarly, in order to test the r$_0$-$M_K$ correlation, the sample is restricted to H$\alpha$ emitters within the region defined by $-24.96<M_K<-21.81$\,(AB) and $41.76<\rm log\,L_{\rm H\alpha}<42.22$ erg\,s$^{-1}$ (see Figure \ref{matchMK_Ha}) in which sub-samples split on the basis of their $M_K$ have the same $L_{\rm H\alpha}$ distributions (means of $41.97$, 41.95, 41.95 and 41.97 and standard deviations of $0.12$ in all instances; ordered from highest to lowest in $M_K$) -- c.f. Table 2. The angular correlation functions of these matched sub-samples are then computed and values of r$_0$ derived as fully described before.

The results are presented in Table 2 and in Figures \ref{r_0_Lalpha} and \ref{r_0_dep}. These reveal that both H$\alpha$ luminosity and rest-frame $K$ luminosity are relevant for the clustering of star-forming galaxies, since for a fixed distribution of one of these, a variation in the other one leads to a change in r$_0$. However, it should be noted that the correlation between r$_0$ and H$\alpha$ luminosity (for a fixed $M_K$ distribution) is maintained as highly statistically significant, while the statistical significance of the correlation between r$_0$ and $M_K$ at a fixed $L_{\rm H\alpha}$ distribution is slightly reduced.

\section{The Clustering of H$\alpha$ emitters across cosmic time} \label{lf_morf}

The clustering properties of the H$\alpha$ emitters at $z=0.84$ can be compared with those of the HiZELS sample at $z=2.23$ (G08) and with results derived from lower redshift surveys of H$\alpha$ emitters at $z=0.24$ and $z=0.4$ \citep{Shioya2008,Nakajima}. This can then be interpreted in the context of the $\Lambda$CDM cosmological model.

\subsection{The clustering of H$\alpha$ emitters at $\bf z=0.24$} \label{clust024}

With the motivation of reliably comparing the clustering measurements at different redshifts, and since the \cite{Shioya2008} sample is publicly available, $\omega(\theta)$ is carefully re-computed for this sample.

\cite{Shioya2008} used galaxy colours to identify $\sim1000$ candidate $z=0.24$ H$\alpha$ emitters within their narrow-band excess sample. For this analysis, the sample is restricted to 492 sources by limiting it to a very robust completeness limit (L$_{\rm H\alpha}$\,$>10^{39.8}$\,erg\,s$^{-1}$, corresponding to SFR $> 0.05$\,M$_{\odot}$\,yr$^{-1}$). The robustness of this sample can be tested and improved taking advantage of a large set of spectroscopic data recently made available in the COSMOS field by the $z$COSMOS project. There are 75 H$\alpha$ candidates with a spectroscopic redshift available from $z$COSMOS and 69 are identified as H$\alpha$ emitters at $0.236<z<0.252$. The remaining 6 sources are identified as [S{\sc ii}]\,6717 emitters  at $0.229<z<0.231$ \citep[contamination by S{\sc ii} emitters is even more significant at lower line fluxes, as pointed out recently by][]{Westra}. With the robust luminosity cut used here, this corresponds to a contamination of 8 per cent, and by removing the 6 confirmed contaminants from the sample of H$\alpha$ emitters, the contamination is estimated to be 7 per cent for the remaining sample of 486 emitters. This results in underestimating r$_0$ by a maximum of 8.4 per cent and r$_0$ will be corrected by 7 per cent to account for contamination in the sample at $z=0.24$; this is 80 per cent of the maximum correction, consistent with the approach used for $z=0.84$.

Finally, in order to obtain r$_0$ (following the same procedures as for the $z=0.84$ sample, except AGN contamination), two redshift distributions are considered: the one assumed in Shioya et al. (2008; a top-hat characterized by $z=0.242\pm0.009$), and a redshift distribution derived from the distribution of the 69 spectroscopically confirmed H$\alpha$ emitters, which can be well-described by a Gaussian with an average of $z=0.245$ and a standard deviation of $\sigma=0.006$. In practice, the values derived from both distributions agree well, but for consistency in determining r$_0$, the Gaussian distribution will be assumed.

\begin{figure*}
\begin{minipage}[b]{0.48\linewidth} 
\centering
\includegraphics[width=7.88cm]{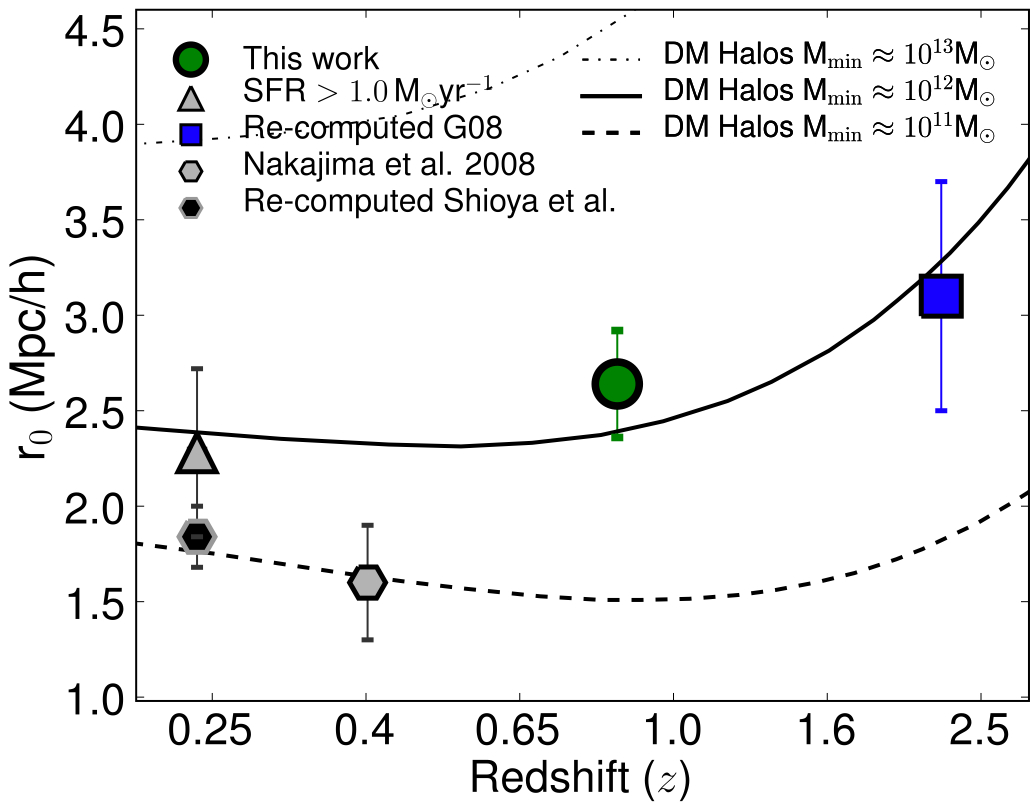}
\end{minipage}
\hspace{0.1cm} 
\begin{minipage}[b]{0.48\linewidth}
\centering
\includegraphics[width=8.30cm]{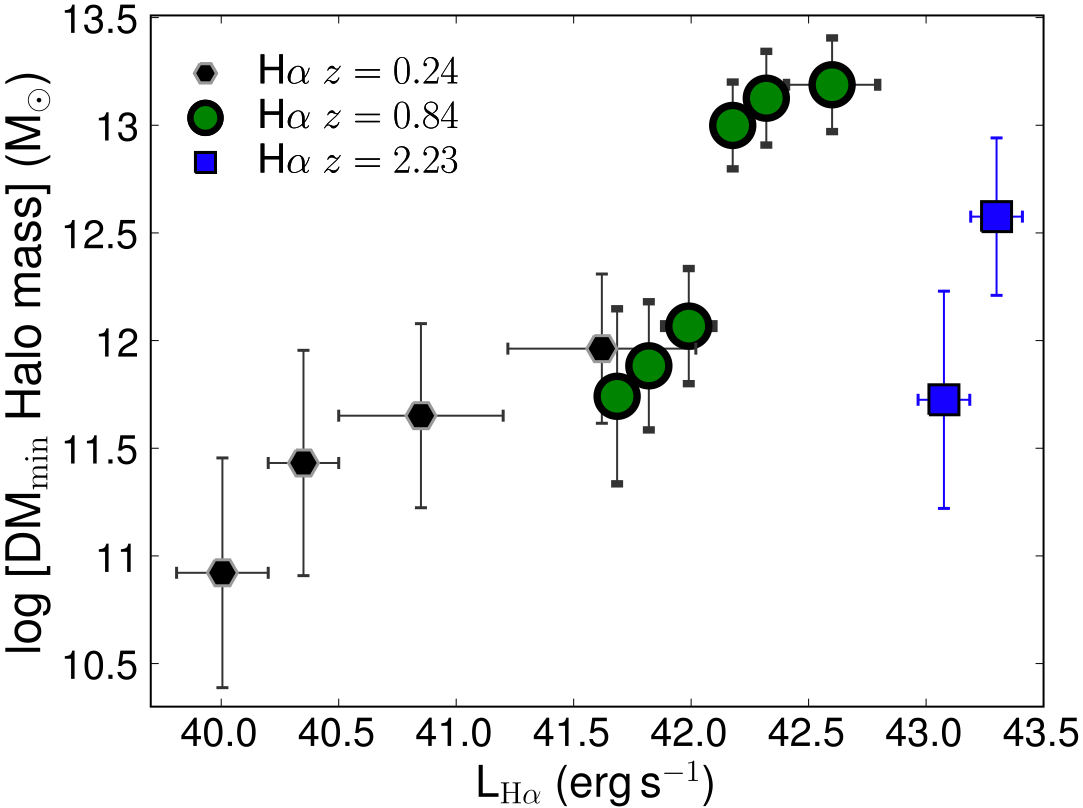}
\end{minipage}
\caption[r0 evo]{$Left$: The clustering length (r$_0$) as a function of redshift for H$\alpha$ emitters selected by narrow-band surveys. This reveals that the H$\alpha$ emitters at $z=0.84$ studied by HiZELS reside in typical dark matter haloes of M$_{\rm min}$\,$\approx10^{12}$\,M$_{\odot}$, consistent with being the progenitors of Milky-Way type galaxies. The lower luminosity H$\alpha$ emitters found in smaller volumes at $z=0.24$ and $z=0.4$ reside in less massive haloes; however, the most luminous H$\alpha$ emitters at $z=0.24$ cluster more strongly, and seem to reside in M$_{\rm min}$\,$\approx10^{12}$\,M$_{\odot}$, while H$\alpha$ emitters at $z=2.23$ reside in haloes just slightly less massive. $Right$: The minimum mass of host dark matter haloes as a function of H$\alpha$ luminosity; this reveals that more luminous emitters reside in more massive haloes, at any given cosmic time, but it also shows that the relation between halo mass and H$\alpha$ luminosity/SFR evolves across cosmic time, with M$_{\rm min}$\,$\approx10^{12}$\,M$_{\odot}$ being much more effective (in respect to SFRs) at $z\sim2$ than at $z<1$.  \label{r_0_evo}}
\end{figure*}

The correlation length results in r$_0=1.8\pm0.2$\,h$^{-1}$\,Mpc, with the error including random errors (from the standard deviation obtained after 1000 measurements of $\omega(\theta)$) and cosmic variance (assumed to affect r$_0$ by 11 per cent for this area). Only separations of \,$5<\theta<800$ arcsec are used when obtaining a power-law fit: angular separations lower than 5\,arcsec are not used as $\omega(\theta)$ becomes steeper, deviating from the power law; this is simply interpreted as the transition between the two-halo and one-halo clustering, happening at $\approx20$\,kpc. It should also be noted that the Limber equation works well for this case, with the differences between using such approximation and integrating the full relation between $\omega(\theta)$ and $\xi(\rm r)$ being lower than 5 per cent for $\theta<800$\,arcsec.

\subsection{The clustering of H$\alpha$ emitters at $\bf z=2.23$} \label{clust223}

It has been mentioned that Limber's equation breaks down for large angular separations and for very narrow filters. Indeed, at $z=2.23$ the co-moving space width of the filter is extremely narrow, presenting $\sigma=7.3$\,h$^{-1}$\,Mpc, and representing only $\approx0.2$ per cent of the total co-moving distance to $z=2.23$; this means that if the spatial correlation function of the H$\alpha$ emitters is a power-law $\xi=(\rm r/r_0)^{-1.8}$, then $\omega(\theta)$, measured at $5<\theta<1000$\,arcsec can not be well-fitted by a power-law with $\beta=-0.8$, since this regime is probing the exact change between $\beta=\gamma+1$ and $\beta=\gamma$.

The correlation length r$_0$ is therefore re-computed. This is done by using the $\omega(\theta)$ data points presented in G08 and by doing a $\chi^2$ fit to the obtained $\omega(\theta)$ from Equation 3. This assumes that the spatial correlation function is a power-law with $\gamma=-1.8$ (which is shown to reproduce the data very well) and only r$_0$ is allowed to vary. This procedure results in r$_0=2.6\pm0.5$\,h$^{-1}$\,Mpc, with the error being derived directly from the $\chi^2$ fit. Note that this implies a best fit $\omega(\theta)$ with $\beta\approx-1$ (for $\gamma=-1.8$) over the separations studied, agreeing well with the G08 best $\beta$ fit to $\omega(\theta)$ of $\beta=-1$. The value of r$_0$ is notably lower than the value of r$_0=3.6\pm0.4$\,h$^{-1}$\,Mpc derived by G08 from using the Limber approximation and their fitted values for $A$ and $\beta$. A further correction is applied to account for a 15 per cent known contamination (the confirmed contaminants are all at different redshifts; this results in underestimating r$_0$ by 20 per cent) based on limited spectroscopic data available at the moment (J. Geach et al. in prep.); this finally results in r$_0=3.1\pm0.7$\,h$^{-1}$\,Mpc (this also assumes a 14 per cent error in r$_0$ due to cosmic variance based on the area of the survey, but such uncertainty is likely to be under-estimated).

\subsection{The clustering evolution of H$\alpha$ emitters since $\bf z=2.23$} \label{lf_morf}

Figure \ref{r_0_evo} presents r$_0$ as a function of redshift for H$\alpha$ emitters; this is the first combination of self-consistent clustering measurements for H$\alpha$ emitters spanning more than half of the history of the Universe ($\approx 8$\,Gyrs) whilst probing 4 different well-defined epochs. For comparison, the r$_0(z)$ predictions for dark matter haloes with a fixed $minimum$ mass of M$_{\rm min}>10^{11-13}$\,M$_{\odot}$  \citep{Matarrese97,Moscardini98} are also shown. A $\Lambda$CDM cosmology is assumed, together with an evolving bias model $b(z)$ from \cite{Moscardini98} and the values from that study are used for various fixed minimum mass haloes (c.f. G08).

The results show that H$\alpha$ emitters found in the HiZELS survey at $z=0.84$ (SFRs\,$>3$M$_{\odot}$\,yr$^{-1}$) reside in Milky-Way type haloes of M$_{\rm min}\approx10^{12}$\,M$_{\odot}$ -- the typical haloes where $L^*$ galaxies in the local Universe reside. This contrasts with the low-H$\alpha$-luminosity galaxies for the samples at $z=0.24$ and $z=0.4$ (presenting SFRs\,$>0.05$\,M$_{\odot}$\,yr$^{-1}$). These seem to reside in much less massive haloes (M$_{\rm min}$\,$\approx10^{11}$\,M$_{\odot}$). These low redshift emitters also present very low stellar masses and low luminosities in all available bands and are therefore likely to be small, young, dwarf-like galaxies, very different from the already much more massive and active star-forming galaxies found at $z=0.84$.

Nevertheless, motivated by the correlation between r$_0$ and L$_{\rm H\alpha}$ found both at $z=0.24$ and $z=0.84$, it is found that if one considers only the $\approx5$\, per cent most luminous emitters (in H$\alpha$) at $z=0.24$ (a rough match in L$_{\rm H\alpha}$ to the $z=0.84$ sample), then these are much more clustered than the complete sample, presenting r$_0\approx2.4$\,h$^{-1}$\,Mpc; this is consistent with these emitters residing in dark matter haloes of M$_{\rm min}$\,$\approx10^{12}$\,M$_{\odot}$. These emitters also present stellar masses closer to those presented by the $z=0.84$ H$\alpha$-selected population, suggesting that the brightest $z=0.24$ H$\alpha$ emitters and part of the sample at $z=0.84$ might be related.

At higher redshift ($z=2.23$), one finds H$\alpha$ emitters residing in the dark matter haloes around M$_{\rm min}$\,$\approx10^{12}$\,M$_{\odot}$, and it is therefore possible that at least a fraction of the very actively star-forming H$\alpha$ emitters found at $z=2.23$ (SFRs $>70$\,M$_{\odot}$\,yr$^{-1}$) by G08 with HiZELS will turn into the typical H$\alpha$ emitters at $z=0.84$ with a strong $L_{\rm H\alpha}^*$ decrease.

\subsection{The dark matter host halo-$\bf L_{\rm \bf H\alpha}^*$ relation} \label{lf_morf}

The right panel of Figure \ref{r_0_evo} presents how the minimum mass of the host dark matter halo changes with measured H$\alpha$ luminosity (all luminosities are derived assuming a constant 1 mag of extinction as in S09). This shows that while the host halo mass increases with luminosity at any given redshift, there seems to be a different relation for each redshift/epoch. For a given dark matter halo mass, one finds that star-formation is tremendously more effective at high-redshift than at lower-redshit; this difference is especially pronounced from $z=0.84$ to $z=2.23$. On the other hand, G08 and S09 demonstrated that there is a clear evolution in the H$\alpha$ luminosity function, showing that the characteristic luminosity, $L_{\rm H\alpha}^*$, evolves by a factor of $>20$ from the local Universe to $z=2.23$; the $L_{\rm H\alpha}^*$ evolution is also most pronounced from $z=0.84$ to $z=2.23$. This suggests that there could be a relation between the host dark matter halo and $L_{\rm H\alpha}^*$ found at different epochs.

\begin{figure}
\centering
\includegraphics[width=8.2cm]{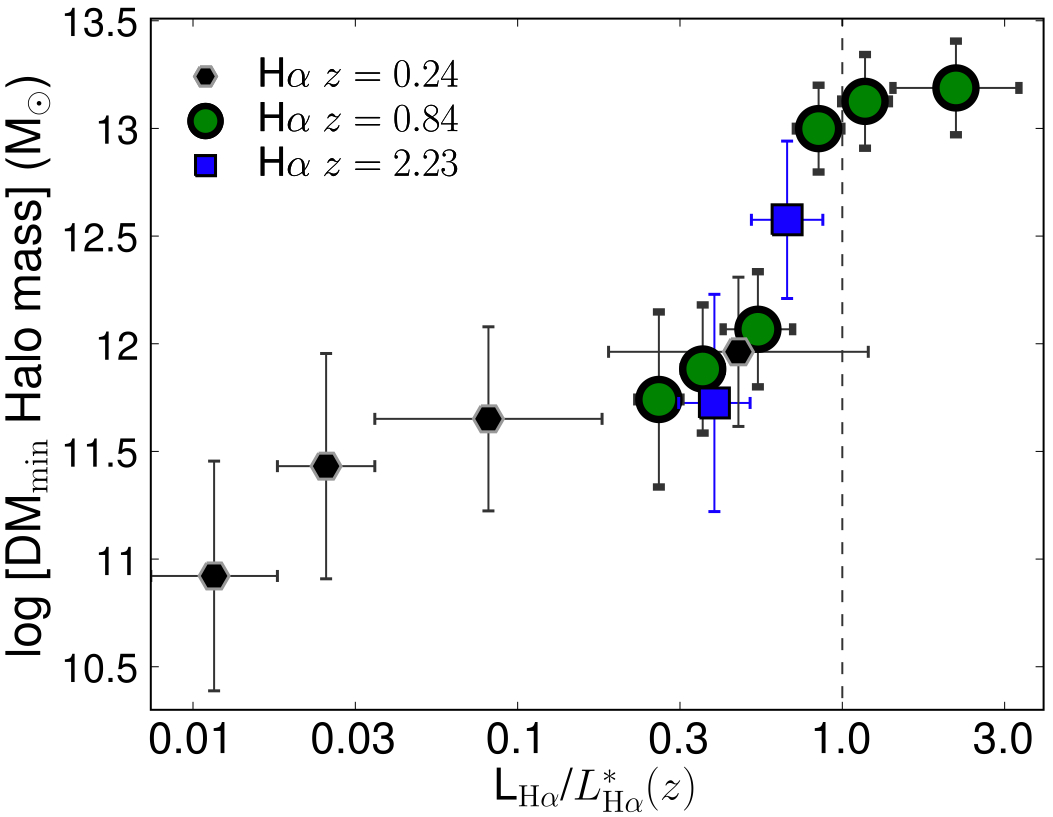}
\caption[wteta_subsam]{The minimum mass of host dark matter haloes as a function of L$_{\rm H\alpha}$/$L_{\rm H\alpha}^*(z)$ ($L_{\rm H\alpha}^*(z)$ is the break in the H$\alpha$ luminosity function, shown to evolve greatly from the local Universe up to $z=2.23$ -- see S09 for more details). This reveals a good agreement between the data at 3 very distinct epochs of the history of the Universe. This shows that luminous H$\alpha$ emitters reside in more massive haloes, but also that across 8 billion years, emitters at the same fraction of $L_{\rm H\alpha}^*$ seem to reside in haloes of typically the same mass.  \label{Halo_L*}}
\end{figure}

In order to investigate this, the results from the right panel of Figure
\ref{r_0_evo} are shown in Figure \ref{Halo_L*} after scaling the measured
luminosities by $L_{\rm H\alpha}^*$ found at each individual redshift\footnote{The
best-fit values for $L_{\rm H\alpha}^*$ are taken from Shioya et al. (2008) at
$z=0.24$, from S09 at $z=0.84$, and from the recent
determination of \cite{Hayes}
at $z=2.23$. The Hayes et al.\ measurement combines the results of G08 with deeper
HAWK-I observations, and suggests a $\sim 0.2$\,dex higher value of
$L_{\rm H\alpha}^*$ than G08 derived, but within the errors of that study.}.
Remarkably, this scaling reveals a clear relation between the typical dark matter
halo host of H$\alpha$ emitters and the value of L$_{\rm H\alpha}$/$L_{\rm H\alpha}^*$,
removing essentially all of the evolution seen in Figure \ref{r_0_evo}. Even though
there is a significant evolution in the H$\alpha$ emitters population from $z=2.23$
to the present epoch, at $L_{\rm H\alpha}^*$, H$\alpha$ emitters seem to reside in
M$_{\rm min}$\,$\approx10^{13}$\,M$_{\odot}$ dark matter haloes, independently of cosmic time. Galaxies below the luminosity function break reside in least massive haloes, while (at least for $z=0.84$), H$\alpha$ emitters above $L_{\rm H\alpha}^*$ seem to reside in haloes just slightly more massive than M\,$\approx10^{13}$\,M$_{\odot}$.

As these results suggest an epoch-independent constancy between $L_{\rm H\alpha}^*$ and the minimum mass of the dark matter halo host of H$\alpha$ emitters, it seems plausible to suggest a simple connection between the two properties. Indeed, it is possible that the strong evolution in the break of the H$\alpha$ luminosity function is being driven by the quenching of star-formation for galaxies residing in haloes much more massive than M\,$\approx10^{12}$\,M$_{\odot}$, either because such haloes favour very intense and fast bursts of star-formation, capable of using the majority of the gas, or because such halo masses create the necessary conditions for AGN feedback to become important.

\section{Conclusions}\label{sum_clustering}

The clustering properties of H$\alpha$ emitters at $z=0.84$ have been fully detailed. These emitters are moderately clustered, with their angular correlation function $\omega(\theta)$ being very well fitted by a power law $A\theta^{-\beta}$ with $A=15.5\pm4.1$ (for $\theta$ in arcsec) and $\beta=-0.81\pm0.05$ out to 600 arcsec (4.5 Mpc). The exact relation between the spatial and angular correlation funtion is used to show that the H$\alpha$ emitters at $z=0.84$ present r$_0=2.7\pm0.3$\,h$^{-1}$\,Mpc for a spatial correlation function given by $\xi=(\rm r/\rm r_0)^{-1.8}$, with the errors accounting for cosmic variance.

A strong dependence of the correlation length on H$\alpha$ luminosity is found, with the most actively star-forming galaxies presenting $5.2\pm1.0$\,h$^{-1}$\,Mpc, while the lower H$\alpha$ luminosity galaxies present a spatial correlation as low as $2.1\pm0.6$\,h$^{-1}$\,Mpc. The correlation length also depends on rest-frame $K$ luminosity (broadly tracing stellar mass) for star-forming galaxies at $z\sim1$ but not on rest-frame $B$ luminosity (or only very weakly).

The r$_0$ correlation with L$_{\rm H\alpha}$ is fully maintained at a fixed rest-frame $K$ luminosity, clearly revealing that it is not a simple result of the L$_{\rm H\alpha}$-$M_K$ correlation; the r$_0$-$M_K$ correlation is also maintained at a fixed L$_{\rm H\alpha}$ distribution, but slightly weakened.

Irregular galaxies and mergers are found to cluster more strongly than discs and non-mergers, respectively. This, however, seems to be driven by the different H$\alpha$ luminosity and $M_K$ distributions of the distinct populations; once they are matched on the two properties, irregulars and discs present the same r$_0$ and mergers and non-mergers are consistent within $\approx1\sigma$. Bulge-to-disc fraction is also shown not to be important to the clustering of star-forming galaxies at $z\sim1$, and morphology seems to be as unimportant for the clustering at high redshift as it is in the local Universe.

H$\alpha$ emitters at $z=0.84$ found by HiZELS reside in minimum dark matter haloes of $\approx10^{12}$\,M$_{\odot}$ similar to those of Milky Way type galaxies in the local Universe. Those are roughly the same as the haloes hosting the brightest H$\alpha$ emitters at $z=0.24$, and just slightly higher mass haloes than the hosts of H$\alpha$ emitters at $z=2.23$. Furthermore, the minimum dark matter halo mass hosting H$\alpha$ emitters increases with H$\alpha$ luminosity, and a $L_{\rm H\alpha}^*$ scaling is able to combine observational results probing the last 8 billion years of the age of the Universe in one single relation. This also suggests a connection between the strong $L_{\rm H\alpha}^*$ evolution of the H$\alpha$ luminosity function and star-formation being truncated in galaxies residing within dark matter haloes with masses much higher than $\approx10^{12}$\,M$_{\odot}$.

The results presented in this study provide important details on the clustering and evolution of H$\alpha$-selected star-forming galaxies at $z=0.84$, which are mostly discs, but with a significant population of irregular and mergers above $L_{\rm H\alpha}^*$. Besides identifying and quantifying clustering relations with fundamental galaxy observables at $z\sim1$ for the first time, the results clearly show that the highest star-formation rates at any epoch only occur in galaxies residing within massive haloes. Dark matter haloes of a given mass seem to be more effective at providing the conditions for intense star-bursts at higher redshift. On the other hand, the fact that the brightest H$\alpha$ emitters become rarer with cosmic time down to the present (seen by the strong evolution in the H$\alpha$ luminosity function; e.g. G08, S09) suggests that the most massive haloes not only provide the conditions and the environment required for the highest SFRs to take place, but they also seem to be the sites in which the quenching of star-formation in galaxies occurs across cosmic time.

\section*{Acknowledgments}

The authors thank the anonymous referee for the prompt and useful comments. DS acknowledges the Funda{\c c}{\~ao para a Ci{\^e}ncia e Tecnologia (FCT) for a doctoral fellowship. PNB and TSG acknowledge support from the Leverhulme Trust. JEG \& IRS thank the U.K. Science and Technology Facility Council (STFC). JK thanks the DFG for support via the German-Israeli Project Cooperation grant GE625-15/1. The authors fully acknowledge the crucial role and unique capabilities of UKIRT and its staff in delivering the extremely high-quality data without which this paper -- and all HiZELS science results published so far -- would not be possible. The authors also thank John Peacock, Tom Shanks and Peder Norberg for valuable suggestions and Jim Dunlop and Karina Caputi for helpful discussions.

\bibliographystyle{mn2e.bst}
\bibliography{bibliography.bib}

\appendix

\bsp

\label{lastpage}

\end{document}